\documentclass{article}

\usepackage{theorem,fullpage,latexsym}
\usepackage{epstopdf}

\newtheorem{all}{Proposition}

{\theorembodyfont{\upshape}}
{\theorembodyfont{\upshape}\newtheorem{example}[all]{Example}}

\newenvironment{remark*}{\noindent {\bf Remark:}}{}

\newcommand{\dom}{\mathrm{dom}\,}

\newcommand{\LTL}{\mathrm{LTL}}
\newcommand{\CTL}{\mathrm{CTL}}
\newcommand{\QCTL}{\mathrm{QCTL}}

\newcommand{\ATL}{{\rm ATL}}
\newcommand{\SL}{{\rm SL}}
\newcommand{\ATLSC}{\ATL_{\mathit{sc}}}
\newcommand{\ATLSCPref}{\ATL_{\mathit{sc},<}}

\newcommand{\infin}{\mathrm{inf}}
\newcommand{\fin}{\mathrm{fin}}

\newcommand{\sem}[1]{\|#1\|}
\newcommand{\atlD}[1]{\langle\!\langle#1\rangle\!\rangle}
\newcommand{\atlB}[1]{[\![#1]\!]}
\newcommand{\atlDclm}[1]{\langle\!\cdot#1\cdot\!\rangle}
\newcommand{\atlBclm}[1]{[\!\cdot#1\cdot\!]}
\newcommand{\atlCclm}[1]{\cdot\!\rangle#1\langle\!\cdot}

\def\Next{\mathop{\mbox{\raisebox{1.7pt}{$\scriptstyle\bigcirc$}}}}
\newcommand{\until}[2]{{({#1}{\mathsf U}{#2})}}
\newcommand{\wuntil}[2]{{({#1}{\mathsf W}{#2})}}

\newcommand{\Act}{{\mathit{Act}}}

\newcommand{\Cl}{{\mathit{Cl}}}
\newcommand{\out}{{\mathrm{out}}}

\newcommand{\Ag}{{\Sigma}}

\def\defeq{{\,\hat{=}\,}}
\newcommand{\angles}[1]{\langle{#1}\rangle}
\def\unfoldOne{{\mathbf{1}}}
\def\existsOne{{\mathop{\exists}\limits_{1}}}

\def\existsSim{{\mathop{\exists}\limits_{\sim}}}
\def\forallSim{{\mathop{\forall}\limits_{\sim}}}

\newcommand{\oomit}[1]{}

\newcommand{\bm}[1]{\mathbf{#1}}
\def\unfoldOne{{\mathbf{1}}}

\begin{document}

\title{On Qualitative Preference in Alternating-time Temporal Logic with Strategy Contexts}

\oomit{
\author{
\name{Dimitar P. Guelev\textsuperscript{a}\thanks{CONTACT Dimitar P. Guelev. Email:gelevdp@math.bas.bg}}
\affil{\textsuperscript{a}Department of Algebra and Logic,
  Institute of Mathematics and Informatics, Bulgarian Academy of Sciences,
  Sofia, Bulgaria}
}}

\author{Dimitar P. Guelev\\
Institute of Mathematics and Informatics,\\
Bulgarian Academy of Sciences\\
{\tt  http://www.math.bas.bg/\~{}gelevdp}
}

\maketitle

\begin{abstract}
We show how to add and eliminate binary preference on plays in Alternating-time Temporal Logic with Strategy Contexts ($\ATLSC^*$) on Concurrent Game Models (CGMs) by means of a translation which preserves satisfaction in models where preference-indiscernibility between plays is an equivalence relation of finite index. The elimination technique also works for a companion second-order path quantifier, which makes quantified path variables range over sets of plays that are closed under preference-indiscernibility. We argue that the preference operator and the specialized quantifier facilitate formulating interesting solution concepts such as Nash equilibrium and secure equilibrium in a straightforward way. We also present a novel translation from $\ATLSC^*$ to Quantified Computation Tree Logic ($\QCTL^*$). Together with the translation which eliminates preference and the specialized form of quantification, this translation allows reasoning about infinite multiplayer synchronous games on CGMs to be translated from the proposed extension of $\ATLSC^*$ into $\QCTL^*$. The setting is related to that of ordered objectives in the works of Bouyer, Brenguier, Markey and Ummels, except that our focus is on the use of the temporal logic languages mentioned above, and we rely on translations into $\QCTL^*$ for the algorithmic solutions.

\end{abstract}

\section*{Introduction}
\label{introduction-sec}

Reasoning about infinite multiplayer games is the intended application of temporal logics of strategic ability. Solving a game means finding the strategies that players can be expected to adopt with their various objectives in mind. Players are assumed to choose their actions rationally; the inherent ambiguity of this assumption and the diversity of games have given rise to a variety of {\em solution concepts}. A solution concept is a collection of requirements that strategy profiles are supposed to meet, in order to be regarded as `optimal' and `rational'. Interesting solution concepts include
{\em Nash equilibrium}, {\em $k$-resilient} Nash equilibrium, which is a generalization of Nash equilibrium about up to $k$ defectors \cite{DBLP:conf/gamesec/Halpern11}, {\em secure equilibrium}, where it is additionally required that a player can not harm other players by deviating from the equilibrium profile \cite{DBLP:journals/tcs/ChatterjeeHJ06}, {\em dominant strategies}, {\em subgame-perfect equilibrium}, and others. Solving games wrt such concepts is the topic of {\em rational synthesis}, which was first considered in a temporal setting in \cite{DBLP:conf/tacas/FismanKL10}. The fitness of a logic for reasoning about any particular class of games greatly depends on how well it can express solution concepts. 

Apart from monadic second order predicate theories such as that of \cite{DBLP:journals/corr/abs-1907-09100}, there are two major groups of logics of strategic ability in the literature. {\em Alternating-time Temporal Logics} ($\ATL$s, \cite{AHK97,AHK02}) can be viewed as generalizations of Computation Tree Logics ($\CTL$s), where the path (branching) quantifier is replaced by a game-theoretic modality: the informal reading of $\atlD{\Gamma}B$, where $\Gamma$ is a coalition of players and $B$ is a temporal property, is {\em coalition $\Gamma$ can enforce $B$ regardless of what the rest of the players do}. Time, and, even more importantly, strategies, are implicit in $\ATL$s: no dedicated symbols such as variables for time instants, plays or strategies are present in the language. Unlike that {\em Strategy Logics} ($\SL$s, \cite{DBLP:conf/fsttcs/MogaveroMV10}, see also the variant \cite{ChatterjeeHP10}) have individual variables that denote strategies and are subject to quantification as in predicate logic, and there is a dedicated construct to express whether the play that results from implementing a given global strategy profile satisfies a given temporal property. Concurrent Game Models (CGMs) possibly with some additional component to represent, e.g., epistemic indistinguishability, and various forms of {\em interpreted systems} are the established semantics of both $\ATL$s and strategy logics. The key component of both classes of structures is the {\em outcome function} which defines the correspondence between players' actions and successor states in plays. Unlike that, in \cite{DBLP:journals/iandc/GutierrezHW15} and its sequel studies, infinite multiplayer games with $\LTL$ objectives and preference on them have been investigated in a notation where every player is in possession of some of the propositional variables from the propositional $\LTL$ vocabulary for objectives. The correspondence between the various notations and the related possibility to port results are beyond the scope of this paper. We only note that the ultimate expressive power of this notation is the same as that of CGMs and therefore, as far as model checking is concerned, any differences are limited to affecting only computational complexity.

Interesting solution concepts are known to be expressible in logics from both groups. This has been demonstrated on the example of Nash equilibrium in \cite{DBLP:conf/fsttcs/MogaveroMV10} and \cite{LopesLM10} for $\SL$ and the extension of $\ATL$ known as $\ATL$ {\em with strategy contexts}, respectively. Interestingly, in both cases the expression is achieved without extending the logics to enable specifying {\em preference}. Reasoning about Nash equilibrium using the proposed formulas focuses on the existence of an equilibrium that achieves some particular tuple of objectives, one for every player. The formulas do not refer to the existence of other objectives that are more or less preferable than the target ones. Therefore reducing the problem of finding an equilibrium to model checking the given formulas in a situation where players may each have a number of objectives with various degrees of desirability requires some non-trivial reasoning {\em outside the logic}. In general this suggests that, when solving games wrt a solution concept with $\ATLSC^*$ or $\SL$, establishing just the soundness of the encoding of that concept by a logical formula is non-trivial. The obvious way for fitting the complete picture of a game where players have multiple objectives and solutions depend on the relative desirability of these objectives is to consider an extension by a preference operator, and this is the main topic of this paper. We adopt the binary operator $<$ for preference that is known from (non-temporal) modal logic \cite{DBLP:conf/kramas/DegremontK08,vanOtterloo2005preference}. Given some conditions $B_1$ and $B_2$ on plays, $B_1<B_2$ means that the plays satisfying $B_1$ are worse than those satisfying $B_2$. The precise meaning of this depends on whether the comparison is supposed to hold between {\em all} the $B_1$-plays and {\em all} the $B_2$-plays, or between {\em some} $B_1$-play and all the $B_2$-plays, etc. Our study focuses on the properties of $<$ in a temporal setting, and on a way to finitely describe the underlying relation which enables algorithmic support. Systems of temporal logic with preference have been studied in the literature from other viewpoints, cf. e.g., \cite{DBLP:journals/jolli/BaskentM20}. 

In this paper we show how $<$ works in Alternating-time Temporal Logic with Strategy Contexts $\ATLSC^*$ as known from \cite{DBLP:conf/fsttcs/MogaveroMV10}. Strategy contexts have been added to systems of $\ATL$ in several variants \cite{WaltherHW07,LopesLM10,WangHY11}. Our technical results apply to the class of models where the binary relation $<$ of preference on plays admits finite description as follows. Given a player $i$, plays $\bm{w}'$ and $\bm{w}''$ are {\em preference-indistinguishable} to $i$, if, for any play $\bm{v}$, $\bm{w}'$ and $\bm{w}''$ are either both better than $\bm{v}$, or both worse than $\bm{v}$, or both incomparable to $\bm{v}$ from the viewpoint of $i$. Our key assumption is that preference indistinguishability is an equivalence relation of finite index, that is, it partitions the set of all plays into finitely many equivalence classes. Furthermore, we assume that the respective equivalence classes are definable by $\LTL$ formulas in the vocabulary of the considered model. Given this assumption, the preferences of player $i$ can be completely described using $i$'s corresponding preference operator $<_i$ on the $\LTL$ formulas which define the classes of preference-indistinguishable plays. These formulas are natural to view as $i$'s objectives. 

Nash equilibrium in multiplayer games with finite systems of temporal objectives and preference on them specified by partial orders has been thoroughly studied in \cite{DBLP:conf/fossacs/BouyerBMU12,DBLP:journals/corr/BouyerBMU15}, with no emphasis on the use of temporal languages. Among other things, this study includes a variety of ways to introduce partial order preference relations built starting from a given finite set of primitive goals, and preference relations satisfy our assumption on definability. The main result about $\ATLSC^*$ with preference in this paper is that the above assumption enables the {\em elimination} of the preference operator. The benefit of this is that some really straightforward expressions of game-theoretic properties that become possible by adding preference to $\ATLSC^*$ can be handled using just a satisfaction-preserving translation into $\ATLSC^*$ without $<$, and no dedicated algorithmic support needs to be developed from scratch for the enhanced language. Of course, this comes at a cost: the computational cost cannot be claimed to be optimal. The example of \cite{DBLP:journals/corr/BouyerBMU15} shows how focusing on a particular solution concept enables more efficient algorithms. Optimality is similarly the focus of the study of another solution concept, dominant strategies, in \cite\cite{li2025rationalcapabilityconcurrentgames} too. Unlike that, the efficiency of translations such as what we do hinges on the complexity of the target logic $\QCTL^*$ model checking, and improvements are limited to making translations more succinct and containing the use of those constructs of $\QCTL^*$ which are the most expensive computationally.

Along with the use of $<$ and its elimination, we propose adding a specialized form of variables for sets of plays to $\ATLSC^*$ and quantification over them. These variables are restricted to range over sets of plays which are {\em closed under preference-indiscernibility}. This is a strong restriction. However, our examples below show that it works well in case the considered variables are to appear as the operands of $<$. Furthermore, our assumption on preference indistinguishability makes it possible to eliminate quantification over such variables at no additional cost. A quantifier which binds {\em state} propositional variables to sets that are closed under {\em epistemic} indiscernibility, which is useful for dealing with the need to encode uniform strategies in the case of incomplete information in a dedicated variant of $\QCTL^*$ can be seen in \cite{DBLP:journals/tocl/BerthonMMRV21}.

Philosophical considerations lead to mathematically diverse classes of binary preference relations \cite{HanssonPreferenceLogic}, the commonest ones being {\em total pre-orders}. Objectives are {\em partially ordered} in \cite{DBLP:conf/fossacs/BouyerBMU12,DBLP:journals/corr/BouyerBMU15}. The diversity of preference relations suggests that preference on plays is reasonable to be viewed as an independent parameter of solution concepts. The use of a dedicated connective for preference facilitates uniform reasoning about solution concepts that involve preference without having to commit to a concrete class of preference relations and non-trivial conventions on expressing solutions. As it becomes clear below, our approach facilitates this. It allows the particular algebraic properties of $<$ to be treated as `data'.

As pointed out in \cite{DBLP:conf/fsttcs/MogaveroMV10}, $\ATLSC^*$ can express Nash equilibrium in an ultimate sense. Using binary preference and second-order path quantification the way we propose allows Nash equilibrium to be expressed by an $\ATLSC^*$ formula which is almost literally the common English language definition. The existence of an equilibrium profile where player $i$ achieves objective $B_i$, $i\in\Ag$ can be written as:
\[\atlDclm{\Ag}\bigwedge\limits_{i\in\Ag}(\Next G_i\wedge\forallSim{}_i\mathbf{c}(G_i<_{\exists\forall,i}\mathbf{c}\Rightarrow\atlBclm{i}\Next\neg\mathbf{c})).\]
Here the second-order path quantifier $\forallSim{}_i$ makes $\mathbf{c}$ range over the sets of infinite plays which are closed under indiscernibility wrt the preference relation of player $i$ and have some successor state of the reference one as the starting state (hence the use of $\Next$ in $\atlBclm{i}\Next\neg{\mathbf c}$). The formula says that there is a strategy profile (the equilibrium) for the grand coalition $\Ag$ which achieves $\bigwedge\limits_{i\in\Ag}G_i$, and, for every individual player $i$, no set of plays $\mathbf{c}$, of which the worst ones are all strictly preferable to $i$ over some play satisfying $G_i$, can be achieved by just $i$ deviating from this profile. That is, player $i$ cannot single-handedly avoid any of the worst plays from $G_i$. Observe that this formulation of Nash equilibrium takes in account that plays are only partially ordered in a particular way, which is defined through $<_{\exists\forall,i}$: $A<_{\exists\forall,i}B$ holds, if there exists an $A$-play which is worse than all the $B$-plays. This is one of the variants of $<$. In the sequel we work mostly with $A<_{\forall\forall,i} B$, which holds if $i$ prefers any of the $B$-plays over all the $A$-plays. 

It is just as straightforward to express {\em secure equilibrium} \cite{DBLP:journals/tcs/ChatterjeeHJ06}: the grand coalition $\Ag$ can achieve $\bigwedge\limits_{i\in\Ag}G_i$ in such a way that an individual defector can neither gain something, nor harm some other player: 
\[\atlDclm{\Ag}\bigwedge\limits_{i\in\Ag}\bigg(\Next G_i\wedge\forallSim{}_i\mathbf{c}\bigg(\bigvee\limits_{j\not=i}(\mathbf{c}<_{\exists\forall,j}G_j\vee\mathbf{c}<_{\forall\exists,j}G_j) \Rightarrow\atlBclm{i}\Next\neg\mathbf{c}\bigg)\bigg).\]
The disjunction $\mathbf{c}<_{\exists\forall,j}G_j\vee\mathbf{c}<_{\forall\exists,j}G_j$ states that either $\mathbf{c}$ includes a play that is worse to $j$ than all of the plays from $G_j$, or $G_j$ includes a play that is better to $j$ than all of the plays from $\mathbf{c}$. Player $j$ is at a potential loss in both cases and the formula says that $i$ cannot incur such a loss by single-handedly deviating from the equilibrium strategy profile. 
 
For the sake of simplicity, we do the technicalities about a single preference operator $<$, which can be viewed as the temporal form of $<_{\forall\forall}$ from \cite{DBLP:conf/kramas/DegremontK08,vanOtterloo2005preference}, and the second-order quantifier on sets of paths in the setting of propositionally Quantified Computation Tree Logic $\QCTL^*$. $\QCTL^*$ is an established intermediate notation for translating temporal logics of strategic ability, and the translation from $\ATLSC^*$ to $\QCTL^*$ can be defined so that the preference connective is not affected. The elimination of $<$ can then be done in $\QCTL^*$. Adding $<$ to $\QCTL^*$ takes extending Kripke models by a corresponding preference relation on paths. As it becomes clear in Section \ref{atlWithPreferenceSection}, adding separate $<_i$ for every player $i\in\Ag$ in CGMs requires no further effort. $\SL$ translates into $\QCTL^*$ too \cite{DBLP:journals/iandc/LaroussinieM15} but porting our work to logics from the group of $\SL$ is beyond the scope of this paper. Among other things, we highlight that the assumption on preference we adopt allows all the variants $<_{Q_1Q_2}$, $Q_1,Q_2\in\{\forall,\exists\}$ of $<$ to be expressed in terms of $<_{\forall\forall}$. Our translation from $\QCTL^*$ with preference to $\QCTL^*$, which carries over to the case of $\ATLSC^*$ with preference, is based on several axioms, which are sufficient to derive the equivalence between formulas and their translations, and therefore can be viewed as an axiomatization of $\QCTL^*$ and $\ATLSC^*$ with preference relative to $\QCTL^*$ and $\ATLSC^*$, respectively.

We include a translation of $\ATLSC^*$ into $\QCTL^*$, which, as stated above, extends to include preference and quantification over sets of plays as proposed. This translation is different from the one proposed in \cite{DBLP:conf/concur/LopesLM12}. In particular, it uses state propositional quantification as available in $\QCTL^*$ more economically.  

\paragraph{Structure of the paper} We give the relevant preliminaries in Section \ref{preliminaries}. Section \ref{atlscToqctl} presents our translation from $\ATLSC^*$ into $\QCTL^*$. Section \ref{binPrefsection} is about the binary preference operator in $\QCTL^*$, the axioms about $<$, and how it can be eliminated. Section \ref{pathQuantification} is about the restricted preference-related form of second-order path quantification we propose, and how it can be used in specifications and then eliminated too. Sections \ref{binPrefsection} and \ref{pathQuantification} are written in the context of $\QCTL^*$ for the sake of simplicity. In Section \ref{atlWithPreferenceSection} we return to $\ATLSC^*$. In Section \ref{exampleNash} we show how all the translations proposed the paper work together on the example of Nash equilibrium. Related work is pointed to and discussed throughout the paper. The conclusions section is a summary of the results.

\section{Preliminaries}
\label{preliminaries}

This section is about the known notions and results that are used in the paper. We render them in the uniform notation of the paper.

\subsection{Kripke Structures and Concurrent Game Models}
A {\em Kripke structure} is a tuple $\angles{W,w_I,R}$ where $W\not=\emptyset$ is the set of {\em possible worlds}, also called {\em states}, $w_I\in W$ is the actual world, or initial state, and $R\subseteq W\times W$ is an {\em accessibility (transition)} relation. Adding a valuation $V\subseteq W\times AP$ of a vocabulary of {\em atomic propositions} $AP\defeq\{p,q,\ldots\}$ gives rise to a Kripke {\em model} $M\defeq\angles{W,w_I,R,V}$.

A {\em Concurrent Game Model (CGM)} is a tuple of the form $M\defeq\angles{W,w_I,\angles{\Act_i:i\in\Ag},o,V}$ where $W$, $w_I$ and $V$ are as in Kripke models. The transition relation $R$ is replaced to allow encoding transitions which are the work of a set $\Ag\defeq\{1,\ldots,N\}$ of players. States and the sets of {\em actions} $\Act_i$, $i\in\Ag$, are related by means of the {\em outcome function} $o:W\times\prod\limits_{i\in\Ag}\Act_i\rightarrow W$: given $w\in W$, the players $i\in\Ag$ simultaneously choose some actions $a_i\in\Act_i$ and the next state is $o(w,\angles{a_1,\ldots,a_N})$.

Given $\Gamma\subseteq\Ag$, we write $Act_\Gamma\defeq\prod\limits_{i\in\Gamma}\Act_i$. Tuples of actions $a\in\Act_\Gamma$ are called (partial) {\em moves}; $a\in\Act_\Ag$ are called {\em global} moves. The Kripke model $\angles{W,w_I,R,V}$ that underlies CGM $M$ is defined by
\begin{equation}\label{CGMtoKripke}
R(w,v)\defeq(\exists a\in\Act_\Ag)(o(w,a)=v).
\end{equation}
This definition guarantees that $R$ is {\em serial} (total), that is, with at least one successor to every state. In the sequel we tacitly assume that CGMs are also Kripke models, with $R$ as in (\ref{CGMtoKripke}).

Given a CGM $M$ and $R$ defined as above and a $\bm{w}=\bm{w}^0\ldots\bm{w}^{|\bm{w}|-1}\in W^+$, we write 
\[R^\infin_{M}(\bm{w})\defeq\{\bm{v}\in W^\omega:\bm{v}^0\ldots\bm{v}^{|\bm{w}|-1}=\bm{w},(\forall k<\omega)R(\bm{v}^k,\bm{v}^{k+1})\}.\]
for the set of all the infinite plays in $M$ which are continuations of $\bm{w}$. The set $R^\fin_{M}(\bm{w})\subseteq W^+$ of the {\em finite} continuations of $\bm{w}$ is defined similarly. Sets of {\em initial} plays can be defined as $R^\infin_{M}(\bm{w})$ where $\bm{w}^0=w_I$. $R^\infin_{M}(w_I)$ consists of all initial plays.
In Kripke models $R^\fin_{M}(.)$ and $R^\infin_{M}(.)$ are said to consist of {\em paths} as opposed to {\em plays} in CGMs. {\em Histories} and {\em fullpaths} are common terms for finite and infinite paths, respectively, too. 

\subsection{Storing Latest Moves} 
Any CGM $M$ can be transformed into a CGM $M^\unfoldOne$ where {\em moves} $a\in\Act_\Ag$ are stored in their destination states. Assuming $*$ to be some fresh (dummy) object, we let
\[M^\unfoldOne\defeq\angles{W^\unfoldOne,w^\unfoldOne_I,\angles{\Act_i:i\in\Ag},o^\unfoldOne,V^\unfoldOne}\]
where
\[\begin{array}{l}
W^\unfoldOne\defeq W\times\Act_\Ag\cup\{\angles{w_I,*}\},\ \ w^\unfoldOne_I\defeq \angles{w_I,*},\ 
o^\unfoldOne (\angles{w,b},a)\defeq \angles{o(w,a),a}.
\end{array}\]
The vocabulary of $M^\unfoldOne$ is $AP^\Act\defeq AP\cup\bigcup\limits_{i\in\Ag}\Act_i$. This union is assumed to be disjoint. We let
\[V^\unfoldOne (\angles{w,b},x)\defeq
\left\{\begin{array}{ll} 
b\not=* \wedge b_i=x,\mbox{ if }x\in\Act_i;\\
 V(w,x),\mbox{ if }x\in AP.
\end{array}\right.\]
Given a $\bm{w}=w^0w^1\in R^\infin_{M}(w_I)$, $w^0=w_I$, a sequence $a^1,a^2\in\Act_\Ag$ is supposed to exist such that $w^{k+1}=o(w^k,a^k)$ for all $k<\omega$. The corresponding play $\angles{w_I,*}\angles{w^1,a^1},\ldots\in R^\infin_{M^\unfoldOne}(\angles{w_I,*})$ is about a particular sequence of such moves. $V$ on $\bm{w}\in R^\infin_{M}(w_I)$ equals the restriction of $V^\unfoldOne$ to the $AP$ co-ordinate on $\bm{w}$'s corresponding plays in $M^\unfoldOne$.

In \cite{GJ04}, the transformation of CGMs $M$ into their corresponding $M^\unfoldOne$ was key to proving the equivalence between CGMs and alternating transition systems, which are the models of Alternating-time Temporal Logic ($\ATL$) in \cite{AHK97}. This transformation is used for the translation of $\ATLSC^*$ on CGMs into $\QCTL^*$ on Kripke models because it preserves the information about latest decisions upon moving from $o^\unfoldOne$ to $R^\unfoldOne(\angles{w,a},\angles{w',a'})\defeq(\exists b\in\Act_\Ag)(o^\unfoldOne(\angles{w,a},b)=\angles{w',a'})$, which is equivalent to $w'=o(w,a')$.

\subsection{Unfolding into Tree Models} 
Both Kripke models and CGMs admit {\em unfolding} into equivalent tree ones. Given CGM $M$ as above, 
we denote its tree unfolding by $M^T\defeq\angles{W^T,w_I^T,\angles{\Act_i:i\in\Ag},o^T,V^T}$, where:
\[
W^T \defeq  R^\fin_M(w_I),\ o^T(w^0\ldots w^n, a)\defeq w^0\ldots w^n o(w^n,a),\ V^T(w^0\ldots w^n,p) \defeq V(w^n,p).
\]
Unfolding Kripke models into tree ones is defined with the clause about $o^T$ above replaced by
\[R^T(w^0\ldots w^n,v^0\ldots v^m)\defeq m=n+1\wedge w^0\ldots w^n= v^0\ldots v^n\wedge R(w^n,v^m).\]
Unfolding a CGM and unfolding it as a Kripke model for $R$ is as in (\ref{CGMtoKripke}) are the same.  

As it becomes clear below, on a given CGM $M$, the translation of $\ATLSC^*$ into $\QCTL^*$ refers to the Kripke model which underlies $M^\unfoldOne$, wrt the {\em tree semantics} of $\QCTL^*$. This semantics of $\QCTL^*$ is based on the tree unfolding $M^T$ of its Kripke models $M$. Accordingly, the unfolding $(M^\unfoldOne)^T$ of $M^\unfoldOne$ is used for translating $\ATLSC^*$. 
\oomit{
The tree semantics allows quantified propositional variables to have different values in states $\angles{w^0,*}\angles{w^1,a^1}\ldots\angles{w^n,a^n},\angles{v^0,*}\angles{v^1,b^1}\ldots \angles{v^m,b^m}\in(W^\unfoldOne)^T=R^\fin_{M^\unfoldOne}(\angles{w_I,*})$ of $(M^\unfoldOne)^T$ even if $w^n=v^m$. The same value must be there in the {\em structure semantics} of $\QCTL$ and this makes that semantics unsuitable for satisfaction-preserving translations of $\ATL$s.
}

\subsection{Quantified Computation Tree Logic}
The syntax of $\QCTL^*$ involves state formulas $A$ and path formulas $B$, which can be defined by the BNFs:
\[A\,::=\,\bot\,\mid\,p\,\mid\,A\Rightarrow A\,\mid\,\exists B\,\mid\,\exists p A\ \ \ \ \ 
B\,::=\,A\,\mid\,B\Rightarrow B\,\mid\,\Next B\,\mid\,\until B B\]
where $p$ stands for an atomic proposition from the vocabulary of the considered Kripke models. To accommodate propositional quantification as in the tree semantics, given a Kripke model $M=\angles{W,w_I,R,V}$, satisfaction is defined using $M$'s tree unfolding $M^T$ and has the form $M^T,w\models A$ with $w\in W^T$ for state formulas $A$ and $M^T,\bm{w}\models B$ with $\bm{w}\in R^\infin_{M^T}(w)$ for path formulas $B$. The defining clauses are as follows:
\[\begin{array}{lcl}
M^T,w\not\models\bot \\
M^T,w\models p &\mbox{iff}& V(w,p) \\
M^T,w\models A_1\Rightarrow A_2 &\mbox{iff}& M^T,w\models A_2\mbox{ or } M^T,w\not\models A_1 \\
M^T,w\models\exists B &\mbox{iff}& M^T,\bm{w}\models B\mbox{ for some }\bm{w}\in R^\infin_{M^T}(w)\\
M^T,w\models\exists p A &\mbox{iff}& (M^T)_p^X,w\models A\mbox{ for some }X\subseteq W^T\\
\\
M^T,w^0w^1\ldots\models A &\mbox{iff}& M^T,w^0\models A\\
M^T,\bm{w}\models B_1\Rightarrow B_2 &\mbox{iff}& M^T,\bm{w}\models B_2\mbox{ or } M^T,\bm{w}\not\models B_1 \\
M^T,w^0w^1\ldots\models \Next B &\mbox{iff}& M^T,w^1w^2\ldots\models B\\
M^T,w^0w^1\ldots\models \until {B_1}{B_2} &\mbox{iff}&\mbox{for some }n<\omega,\ M^T,w^nw^{n+1}\ldots\models B_2\\
& & \mbox{and }M^T,w^kw^{k+1}\ldots\models B_1,\ k=0,\ldots,n-1.
\end{array}\]
Here, for any Kripke model $M\defeq\angles{W,w_I,R,V}$ and $p\in AP$, $M_p^X\defeq\angles{W,w_I,R,V_p^X}$ where
$V_p^X(w,q)\defeq V(w,q)$ for $q\not= p$ and $V_p^X(w,p)\defeq w\in X$. A state formula $A$ is valid in $\QCTL^*$, written $\models A$, if $M^T,w_I^T\models A$ for all $M$, provided the vocabulary includes all of $A$'s atomic propositions.

Propositional logic constant $\top$ and connectives $\neg$, $\vee$, $\wedge$ and $\Leftrightarrow$, which are expressible by $\bot$ and $\Rightarrow$, are used as usual. Linear Temporal Logic ($\LTL$) can be viewed as the logic of path formulas $B$ in $\QCTL^*$ with $A$ of the BNF allowed to be only $\bot$ or an atomic proposition, and not any state formula. $\LTL$ abbreviations
\[\Diamond B\defeq \until\top B,\ \Box B\defeq\neg\Diamond\neg B,\mbox{ and }\wuntil {B'}{B''}\defeq \until{B'}{B''}\vee\Box B'\]
apply in $\QCTL^*$. The universal dual $\forall$ of $\exists$ is defined as $\neg\exists\neg$, to be used both as a path quantifier, and as a propositional quantifier. 

Using $M^T$ instead of $M$ above facilitates writing the clause about propositional quantification. All the other clauses would work the same if written about $M$. The decidability of validity in $\QCTL^*$ wrt the tree semantics, stems from the results in \cite{DBLP:conf/ausai/French01,French06}. 
\oomit{
$\QCTL^*$ also has a {\em structure semantics}, where the above clauses apply with $M$ instead of $M^T$ in them. In particular, the clause
\[\begin{array}{lcl}
M,w\models\exists p A &\mbox{iff}& M_p^X,w\models A\mbox{ for some }X\subseteq W
\end{array}\]
forces a quantified variable $p$ to have a value determined by the reference state wherever that state appears along plays, unlike the tree semantics, where states along a play in $M^T$ are ever longer finite plays from the original model $M$, and never recur, which means that a bound variable may have different values at states at different positions along a path in $M^T$, even if these states correspond to the same state in the given $M$. Validity is undecidable in the structure semantics.}
An in-depth up-to-date presentation on the whole topic can be found in \cite{DBLP:journals/corr/LaroussinieM14}. 

\subsection{Guarded Normal Form in Linear Temporal Logic}
\label{LTLGNFs}

$\LTL$ formulas $B$ in general are known to admit the {\em Guarded Normal Form (GNF)}
\[\bigvee\limits_k B_k\wedge \Next B_k'\]
where the {\em guard} ({\em head}) formulas $B_k$ are built using only atomic propositions, $\bot$ and $\Rightarrow$, and form a {\em full system}, which means that $B_k$ are pairwise inconsistent and $\bigvee\limits_k B_k$ is valid. GNFs can be obtained by induction on the construction of the given $B$ using, among other things, the fixpoint expansion $\until{B'}{B''}\Leftrightarrow B''\vee B'\wedge\Next\until{B'}{B''}$. Given a path formula $C$, we assume $\Cl(C)$, the {\em closure} of $C$, to be some fixed set that includes $C$ and the {\em tail} formulas $B_k'$ from some GNF of every formula $B$ in it.
Since the {\em modal depth} $d(B_K')$ of the {\em tail} formulas $B_k'$ can be chosen not to exceed that of the considered $B$ in GNFs, $\Cl(B)$ can be chosen to be finite.   

\subsection{Alternating-time Temporal Logic with Strategy Contexts}
 
In $\ATL$s, {\em strategy contexts} are partial strategy profiles, which appear on the LHS of $\models$. Upon evaluating the game-theoretic modality $\atlDclm{\Gamma}$ of $\ATLSC$, which requires the existence of some appropriate strategies for players $i\in\Gamma$, players $i\not\in\Gamma$ who have strategies for them in the reference strategy context are assumed to be sticking to these strategies. This enables the expressive power to be brought closer to what assignments to strategy-valued variables contribute in $\SL$s. The idea is implemented in several variants. To the best of our knowledge, the earliest ones are \cite{WaltherHW07,BrihayeLLM09,LopesLM10,WangHY11}. In this paper we work with the system of $\ATL$ with strategy profiles known as $\ATLSC^*$ from \cite{LopesLM10}. In the notation of our paper, the definition of $\ATLSC^*$ is as follows. 

State formulas $A$ and path formulas $B$ have the syntax
\[A\,::=\,\bot\,\mid\,p\,\mid\,A\Rightarrow A\,\mid\,\atlDclm{\Gamma} B\,\mid\,\atlCclm{\Gamma} A\ \ \ \ \ 
B\,::=\,A\,\mid\,B\Rightarrow B\,\mid\,\Next B\,\mid\,\until B B\]
where $\Gamma\subseteq \Ag$ is a set of players from the considered CGMs. $\atlDclm{.}$ is known as the {\em game-theoretic} or {\em co-operation} modality in $\ATLSC^*$. The role of $\atlCclm{.}$ is explained below. 

Let $M\defeq \angles{W,w_I,\angles{\Act_i:i\in\Ag},o,V}$ be a CGM with $\Ag$ as the set of players and $AP$ as the vocabulary. A {\em strategy for $i\in\Ag$ starting from $w\in W$} is a function $s:R^\fin_M(w)\rightarrow\Act_i$. A strategy {\em profile} for $\Gamma\subseteq\Ag$ from $w$ is a tuple $\rho\defeq\angles{s_i:i\in\Gamma}$ where $s_i:R^\fin_M(w)\rightarrow\Act_i$ for every $i\in\Gamma$. Equivalently, a strategy profile $\rho$ is a function on $\Gamma$ such that $\rho(i)=s_i$. Hence we can we write $\dom\rho$ for $\Gamma$ and apply set-theoretic notation about functions in general. 

Given strategy profiles $\rho'$ and $\rho$, we write $\rho'\circ\rho$ for the extension $\rho'\cup\rho|_{\Ag\setminus\dom\rho'}$ of $\rho$ by $\rho'$, with the strategies from $\rho'$ overriding those from $\rho$ for $i\in\dom\rho\cap\dom\rho'$.

Given $\bm{w}\defeq w^0\ldots w^{|\bm{w}|-1}\in R^\fin_M(w^0)$ and a strategy $s$ for $i\in\Ag$ from $w^0$, we write $\bm{w}^{-1} s$ for the strategy for $i$ which prescribes the same actions as $s$, but is defined on finite histories that start at $w^{|\bm{w}|-1}$: $(\bm{w}^{-1}s)(\bm{v})\defeq s(w^0\ldots w^{|\bm{w}|-2}\cdot \bm{v})$ for $\bm{v}\in R^\fin_M(w^{|\bm{w}|-1})$. Informally, $\bm{w}^{-1} s$ tells $i$ how to continue playing by $s$, once the properly past part $w^0\ldots w^{|\bm{w}|-2}$ of $\bm{w}$ is no longer observable. The notation extends to profiles: $\bm{w}^{-1}\rho\defeq\angles{\bm{w}^{-1} s_i:i\in\dom\rho}$. This notation can be traced back to Brzozowski derivatives on formal languages \cite{DBLP:journals/jacm/Brzozowski64}.	

The infinite plays which can take place if strategy profile $\rho$ from state $w\in W$ is followed by the players from $\dom\rho$ and the actions of the other players are unrestricted form the set
\[\out(w,\rho)\defeq\{w^0w^1\ldots\in R^\infin_M(w):(\forall k<\omega)(\exists b\in\Act_{\Ag\setminus\dom \rho})
(w^{k+1}=o(w^k,\rho(w^0\ldots w^k)\cup b))\}.\] 
In $\ATLSC^*$, satisfaction has the form $M,\rho,w\models A$ with $w\in W$ and $\rho$ being a strategy profile from state $w$ for state formulas $A$, and $M,\rho,w^0w^1\ldots\models B$ with $w^0w^1\ldots\in R^\infin_M(w^0)$ and $\rho$ being a strategy profile from $w^0$ for path formulas $B$. In both cases $\dom\rho$ can be any subset of $\Ag$ and $\rho$ is called the {\em strategy context}. The empty strategy profile is written $\angles{}$. The clauses are as follows:
\[\begin{array}{lcl}
M,\rho,w\not\models\bot \\
M,\rho,w\models p &\mbox{iff}& V(w,p) \\
M,\rho,w\models A_1\Rightarrow A_2 &\mbox{iff}& M,\rho,w\models A_2\mbox{ or } M,\rho,w\not\models A_1 \\
M,\rho,w\models\atlDclm{\Gamma} B &\mbox{iff}& \mbox{there exists a strategy profile }\rho'\mbox{ for }\Gamma\mbox{ from }w,\\
&&\mbox{such that }M,\rho'\circ\rho,\bm{w}\models B\mbox{ for all }\bm{w}\in\out(w,\rho'\circ\rho)\\
M,\rho,w\models\atlCclm{\Gamma} A &\mbox{iff}& M,\rho|_{\Ag\setminus\Gamma},w\models A \\
\\
M,\rho,w^0w^1\ldots\models A &\mbox{iff}& M,\rho,w^0\models A\\
M,\rho,\bm{w}\models B_1\Rightarrow B_2 &\mbox{iff}& M,\rho,\bm{w}\models B_2\mbox{ or } M,\rho,\bm{w}\not\models B_1 \\
M,\rho,w^0w^1\ldots\models \Next B &\mbox{iff}& M,(w^0)^{-1}\rho,w^1w^2\ldots\models B\\
M,\rho,w^0w^1\ldots\models \until {B_1}{B_2} &\mbox{iff}&\mbox{for some }n<\omega,\ M,(w^0\ldots w^n)^{-1}\rho,w^nw^{n+1}\ldots\models B_2\\
& & \mbox{and }M,(w^0\ldots w^k)^{-1}\rho,w^kw^{k+1}\ldots\models B_1,\ k=0,\ldots,n-1.
\end{array}\]
A (state) formula $A$ is {\em valid} in $\ATLSC^*$, if $M,\angles{},w_I\models A$ in all CGMs $M$ whose propositional vocabulary and set of players include the atomic propositions and the players appearing in $A$.

Informally, $M,\rho,w\models\atlDclm{\Gamma}B$ means that there exists a strategy profile $\rho'$ for the members of $\Gamma$ which, if augmented with whatever strategies are there in the context profile $\rho$ for players outside $\Gamma$, enforces $B$ whatever the players outside $\Gamma$ and the context profile do. $B$ itself is evaluated with the considered strategies $\rho'$ for $\Gamma$ added to the context $\rho$, replacing whatever strategies for these players are already in the context $\rho$ for evaluating $\atlDclm{\Gamma}B$. This means that the context profile generally grows to include strategies for more players upon moving to subformulas of $\atlDclm{.}$-formulas. To control this, $\atlCclm{\Delta}A$ is used for `deleting' the strategies for the members of $\Delta$ from the context before evaluating $A$, thus removing the restriction on the players from $\Delta$ to stick to any particular strategies upon evaluating $\atlDclm{.}$-subformulas of $A$. Unless in the scope of an $\atlCclm{.}$, the $\atlDclm{.}$-subformulas of the $B$ to be enforced must hold in the context $\rho'\circ\rho$. The basic $\ATL$ form $\atlD{\Gamma}$ of the game-theoretic modality can be expressed as $\atlCclm{\Ag}\atlDclm{\Gamma}$, with the $\atlCclm{\Ag}$ serving to clear the context strategy profile. Hence no particular (rational) behaviour on behalf of the non-members of a coalition can be taken in account by $\ATL$'s form of the construct: $\atlD{\Gamma}B$ holds only if the members of $\Gamma$ can enforce $B$ whatever the others do. 

Classical propositional logic derived constant $\top$ and connectives, and $\LTL$ derived connectives are included in $\ATLSC^*$ as usual. The universal dual of $\atlDclm{.}$ is $\atlBclm{\Gamma}B\defeq\neg\atlDclm{\Gamma}\neg B$. Informally, $\atlBclm{\Gamma}B$ means that $\Gamma$ cannot prevent $B$, {\em even if the non-members who have strategies in the context profile stick to these strategies}. 

The $\CTL$ (and $\CTL^*$ and $\QCTL^*$) path quantifier $\exists$ on the Kripke model that underlies a given CGM $M$ can be expressed as $\atlBclm{\emptyset}$, which is equivalent to its basic $\ATL$ form $\atlB{\emptyset}$: the empty coalition can prevent nothing. Hence $\CTL$ immediately embeds already into $\ATL$ without strategy contexts, with or without the ${}^*$.

Much like $\QCTL^*$, $\ATLSC^*$ has a subset $\ATLSC$ where only state formulas are considered, and those are defined by requiring every occurrence of an $\LTL$ operator to be preceded by a $\atlDclm{\Gamma}$ or a $\atlBclm{\Gamma}$. Thanks to the availability of strategy contexts, this subset has the same ultimate expressive power as $\ATLSC^*$ \cite{LopesLM10,DBLP:journals/iandc/LaroussinieM15}, whereas $\CTL$ is strictly less expressive than $\CTL^*$.

\subsection{Preference Relations} 

We work with preference modeled as a binary relation $<$ on plays. 
The commonest binary preference relations are partial or linear orders or pre-orders. The reader is referred to \cite{HanssonPreferenceLogic} for an introduction. We assume preferences to be {\em stable}. This means that preferences become updated according to the progress made on achieving them but remain ultimately unchanged. The precise meaning of this is given in symbols in the opening of Section \ref{binPrefsection} and in Section \ref{qctldefsection}.

\begin{quote}
In the rest of the paper, we tacitly assume $M$ to stand for some Kripke model $\angles{W,w_I,R,V}$, or CGM $\angles{W,w_I,\angles{\Act_i:i\in\Ag},o,V}$, depending on the context. A single preference relation denoted by $<$, resp., a system $\angles{<_i:i\in\Ag}$ of preference relations can be there too, if the extensions of $\QCTL^*$, resp. $\ATLSC^*$ by preference proposed in the following sections of this paper are to be interpreted in $M$.
\end{quote}

\section{Translating $\ATLSC^*$ into $\QCTL^*$}
\label{atlscToqctl}

In this section, given a vocabulary of atomic propositions $AP$, a set of players $\Ag$ and an $\ATLSC^*$ (state) formula $A$ written in these symbols, we define a $\QCTL^*$ formula $t(A)$ such that for every CGM\\
$M\defeq \angles{W,w_I,\angles{\Act_i:i\in\Ag},o,V}$ for $AP$ and $\Ag$
\[M,\angles{},w_I\models A\ \leftrightarrow\ (M^\unfoldOne)^T,(w_I^\unfoldOne)^T\models t(A).\]
Here $\models$ stands for the $\ATLSC^*$ satisfaction relation and the $\QCTL^*$ satisfaction relation on the LHS and the RHS of $\leftrightarrow$, respectively, and the transition relation $R^\unfoldOne$ of $M^\unfoldOne$ is defined from $o^\unfoldOne$ as in (\ref{CGMtoKripke}). The translation below is different from the one given in \cite{DBLP:conf/concur/LopesLM12,DBLP:journals/iandc/LaroussinieM15}.

In the sequel we assume that the meaning of $M,\rho,w\models A$ and $M,\rho,\bm{w}\models B$ is clear for state formulas $A$ and path formulas $B$ from both $\ATLSC^*$ and $\QCTL^*$, with $\rho$ ignored in the latter case. This enables writing the above correspondence between the $\ATLSC^*$ formula $A$ and its $\QCTL^*$ translation $t(A)$ as $M,\angles{},w_I\models A\Leftrightarrow t(A)$.

Recall that the vocabulary of $M^\unfoldOne$ is the supposedly disjoint union $AP^\Act\defeq AP\cup\bigcup\limits_{i\in\Ag}\Act_i$. Since $M,\angles{},w\models A$ is known to be equivalent to $M^\unfoldOne,\angles{},\angles{w,a}\models A$ for all $w\in W$ and $a\in\Act_\Ag$, we only need to prove the equivalence between $A$ and $t(A)$ in $M^\unfoldOne$. The tree semantics of $\QCTL^*$ immediately takes considerations to the tree unfolding $(M^\unfoldOne)^T$ of $M^\unfoldOne$, which is both a CGM and a Kripke model just like $M^\unfoldOne$.

\subsection{Encoding Strategies by State Variables}
\label{encodingStrategies}

In our translation of $\ATLSC^*$ into $\QCTL^*$, strategies are encoded by quantified state variables with appropriate $\QCTL^*$-defined restrictions on their valuations. Let $s:R_{(M^\unfoldOne)^T}^\fin(w)\rightarrow\Act_i$ be a strategy for $i\in\Ag$ in $(M^\unfoldOne)^T$ for plays starting from $w\in (W^\unfoldOne)^T$. Observe that, by the construction of the tree unfolding $(M^\unfoldOne)^T$, state $w\in (W^\unfoldOne)^T=R_{M^\unfoldOne}^\fin(w_I^\unfoldOne)$ is a finite play in $M^\unfoldOne$. Therefore $R_{M^\unfoldOne}^\fin(w)$ is the set of $w$'s finite continuations, which are both initial finite plays in $M^\unfoldOne$ because so is $w$, and states of $(M^\unfoldOne)^T$ that are reachable from $w$. Let $X\subseteq R_{M^\unfoldOne}^\fin(w)\subseteq (W^\unfoldOne)^T$,
\begin{equation}\label{Xdef}
X\defeq\{u^0\ldots u^n\in R_{M^\unfoldOne}^\fin(w):|w|\leq n<\omega, V^\unfoldOne(u^n,s((u^{|w|-1}\ldots u^{n-1})^T))\}
\end{equation}
where $(u^{|w|-1}\ldots u^{n-1})^T$ stands for the finite play
\[u^0\ldots u^{|w|-1}\ \ u^0\ldots u^{|w|-1}u^{|w|}\ \ \ \ldots\ \ \ u^0\ldots u^{|w|-1}u^{|w|}\ldots u^{n-1}\]
in $(M^\unfoldOne)^T$ which corresponds to $u^{|w|-1}\ u^{|w|}\ \ldots\  u^{n-1}\in R_{M^\unfoldOne}^\fin(u^{|w|-1})=R_{M^\unfoldOne}^\fin(w^{|w|-1})$ of $M^\unfoldOne$.

Let $p\in AP$. Then an infinite play $\bm{v}\in R^\infin_{(M^\unfoldOne)^T}(w)$ starting from $w$ is consistent with $s$ iff $((V^\unfoldOne)^T)_p^X(v,p)$ for all of $\bm{v}$'s states $v$, except possibly the first one $w$, because $V^\unfoldOne(u^n,s((u^{|w|-1}\ldots u^{n-1})^T))$ means that $u^n$ is reached by $i$ choosing an action prescribed by $s$. Hence, in $((M^\unfoldOne)^T)_p^X$, $\Next\Box p$ defines the set of plays from $R^\infin_{(M^\unfoldOne)^T}(w)$ which are consistent with $s$. We use this in the translation clause for $\atlDclm{.}$-formulas below. Furthermore, the definition (\ref{Xdef}) of $X$ implies that
\[((M^\unfoldOne)^T)_p^X,w\models\forall\Box\bigg(\bigvee\limits_{a\in\Act_i}\forall\Next(p\Leftrightarrow a)\bigg).\]
This means that, for every state $v$ that is reachable from $w$, there exists a unique $a\in\Act_i$ such that $p$ holds only at the successors of $v$ obtained by $i$ choosing $a$. Obviously this $a$ is the value of $s$ on the play from $w$ to $v$. Conversely, let $p$ satisfy 
\[(M^\unfoldOne)^T,w\models\forall\Box\bigg(\bigvee\limits_{a\in\Act_i}\forall\Next(p\Leftrightarrow a)\bigg).\]
Then
\[X\defeq\{v\in (W^\unfoldOne)^T:v\mbox{ is reachable from }w\mbox{ in }(M^\unfoldOne)^T\mbox{ and }(V^\unfoldOne)^T(w,p)\}\]
admits the definition (\ref{Xdef}) for a unique strategy $s$ for $i$ for plays starting from $w$.

The valuation of state variable $p$ {\em encodes} strategy $s$ in $(M^\unfoldOne)^T$, if the above correspondence is in place. Observe that, unlike \cite{DBLP:conf/concur/LopesLM12,DBLP:journals/iandc/LaroussinieM15}, here actions $a\in\Act_i$ label their {\em destination} states. We first used a destination state-based encoding of actions in a translation from epistemic $\ATL$ into epistemic $\CTL$ in \cite{DBLP:journals/corr/abs-1303-0794}. Among other things, labeling destination states and not {\em source} states enables the encoding of partial and total {\em non-deterministic} strategies $s:R^\fin_M(w)\rightarrow{\mathcal P}(\Act_i)$ and $s:R^\fin_M(w)\rightarrow{\mathcal P}(\Act_i)\setminus\{\emptyset\}$. Such encodings are obtained by appropriately relaxing $\bigvee\limits_{a\in\Act_i}\forall\Next(p\Leftrightarrow a)$, which requires $p$ to correspond to a unique action. Subsets of the set of states that are reachable from a given state $w$ are used to define the working of actions in the original \cite{AHK97} alternating transition system-based semantics of $\ATL$. The role of quantified propositional variables $p$ in the above translation is to define such subsets too.

\subsection{The Translation}
\label{theTranslationATLSCtoQCTL}

Given an $\ATLSC^*$ formula $F$ and some subformula $G$ of its, evaluating $F$ at $M,\angles{},w_I$ requires evaluating $M,\rho,w\models G$ for contexts $\rho$ whose domain $\dom\rho\defeq\{i_1,\ldots,i_n\}\subseteq\Ag$ can be determined from the location of the occurrence of $G$ in $F$. The same applies to evaluating occurrences of subformulas in $(M^\unfoldOne)^T$.  We write $t_{i_1,\ldots,i_n}^{p_1,\ldots,p_n}(G)$ to indicate the propositional variables $p_1,\ldots,p_n$ which encode the strategies from the contexts for $i_1,\ldots,i_n$ where $G$ is to be evaluated in the $\QCTL^*$-translation of $G$. The given $F$ is supposed to be equivalent to $t(F)$ when evaluated wrt the empty context $\angles{}$.
We define $t_{i_1,\ldots,i_n}^{p_1,\ldots,p_n}(.)$ by induction on the construction of occurrences of state subformulas $A$ and path subformulas $B$ in $F$ so that, if $\rho$ is a strategy profile for some $\{i_1,\ldots,i_n\}\defeq\dom\rho\subseteq\Ag$, $w\in (W^\unfoldOne)^T$ and $\bm{w}\in R^\infin_{(M^\unfoldOne)^T}(w)$, and the strategies for $i_1,\ldots,i_n$ from $\rho$ are expressed by means of the variables $p_1,\ldots,p_n$ as described in Section \ref{encodingStrategies}, then
\[(M^\unfoldOne)^T,\rho,w\models A\Leftrightarrow t_{i_1,\ldots,i_n}^{p_1,\ldots,p_n}(A)\mbox{ and }(M^\unfoldOne)^T,\rho,\bm{w}\models B\Leftrightarrow t_{i_1,\ldots,i_n}^{p_1,\ldots,p_n}(B).\]
The defining clauses for state formulas are as follows:
\[\begin{array}{lcl}
t_{i_1,\ldots,i_n}^{p_1,\ldots,p_n}(\bot)&\defeq &\bot\\
t_{i_1,\ldots,i_n}^{p_1,\ldots,p_n}(p)&\defeq &p\\
t_{i_1,\ldots,i_n}^{p_1,\ldots,p_n}(A_1\Rightarrow A_2)&\defeq &t_{i_1,\ldots,i_n}^{p_1,\ldots,p_n}(A_1)\Rightarrow t_{i_1,\ldots,i_n}^{p_1,\ldots,p_n}(A_2)\\
t_{i_1,\ldots,i_n}^{p_1,\ldots,p_n}(\atlDclm{\{i_k,\ldots,i_m\}}B) &\defeq &
\exists q_k\ldots\exists q_m
\left(\begin{array}{l}
\bigwedge\limits_{l=k}^m
\forall\Box\bigg(\bigvee\limits_{a\in\Act_{i_l}}\forall\Next(q_l\Leftrightarrow a)\bigg)\wedge\\
\forall\left(\begin{array}{l}\Next\Box\bigg(\bigwedge\limits_{l=1}^{k-1} p_i\wedge\bigwedge\limits_{l=k}^m q_i\bigg)\Rightarrow\\ t_{i_1,\ldots,i_{k-1},i_k,\ldots,i_m}^{p_1,\ldots,p_{k-1},q_k,\ldots,q_m}(B)\end{array}\right)
\end{array}\right)\\
t_{i_1,\ldots,i_n}^{p_1,\ldots,p_n}(\atlCclm{\{i_k,\ldots,i_m\}}A) &\defeq &t_{i_1,\ldots,i_{k-1}}^{p_1,\ldots,p_{k-1}}(A)
\end{array}\]
In the clause for $\atlDclm{\{i_k,\ldots,i_m\}}B$, $q_k,\ldots,q_m$, which are meant to encode the strategies for $i_k,\ldots,i_m$ that are required to exist for $\atlDclm{\{i_k,\ldots,i_m\}}G$ to hold, are supposed to be fresh state variables. Alternatively, a fixed correspondence between the players $1,\ldots,N$ from $\Ag$ and some state variables $p_1,\ldots,p_N\in AP$ for encoding these players' strategies can be used with the inner quantifications of $p_i$ shadowing the outer ones to reflect that $\rho'$ overrides $\rho$ in $\rho'\circ\rho$ for players in $\dom\rho'\cap\dom\rho$ upon translating $\atlDclm{\Gamma}$-subformulas. To simplify the formulation of this clause, it is assumed that the players who both have strategies for them in the reference context and appear in the designated occurrence of $\atlDclm{.}$ are $i_k,\ldots,i_n$. If there are no such players, then $k=n+1$ is assumed. The clauses for path formulas are trivial:
\[\begin{array}{lcl}
t_{i_1,\ldots,i_n}^{p_1,\ldots,p_n}(B_1\Rightarrow B_2)&\defeq &t_{i_1,\ldots,i_n}^{p_1,\ldots,p_n}(B_1)\Rightarrow t_{i_1,\ldots,i_n}^{p_1,\ldots,p_n}(B_2)\\
t_{i_1,\ldots,i_n}^{p_1,\ldots,p_n}(\Next B)&\defeq &\Next t_{i_1,\ldots,i_n}^{p_1,\ldots,p_n}(B)\\
t_{i_1,\ldots,i_n}^{p_1,\ldots,p_n}(\until{B_1}{B_2})&\defeq &\until{t_{i_1,\ldots,i_n}^{p_1,\ldots,p_n}(B_1)}{t_{i_1,\ldots,i_n}^{p_1,\ldots,p_n}(B_2)}
\end{array}\]
Both the translation from \cite{DBLP:conf/concur/LopesLM12} and the one here use atomic propositions for players' actions, which ties them to the given CGMs. In \cite{DBLP:conf/concur/LopesLM12}, these are bound propositional variables and the outcome function is entirely encoded by a formula to specify the restrictions on them, whereas we assume this information to be supplied by moving from $M$ to $M^\unfoldOne$.

\subsection{This Translation and Translations from \cite{DBLP:conf/concur/LopesLM12,DBLP:journals/iandc/LaroussinieM15}}

The translation of $\ATLSC^*$ into $\QCTL^*$ with the tree semantics given in this section is different from the one given in \cite{DBLP:conf/concur/LopesLM12,DBLP:journals/iandc/LaroussinieM15}. It requires first moving from the given CGM $M$ to its corresponding $M^\unfoldOne$ and $|W^\unfoldOne|=1+|W|\prod\limits_{i\in\Ag}|\Act_i|$. However, translating an occurrence of $\atlDclm{\Gamma}$ takes $|\Gamma|$ quantifications and not $\sum\limits_{i\in\Gamma}|\Act_i|$ quantifications as in \cite{DBLP:conf/concur/LopesLM12}. 
A translation which takes $|\Gamma|$ quantifications for every occurrence of $\atlDclm{\Gamma}$ is given for {\em turn-based} games in \cite{DBLP:journals/iandc/LaroussinieM15}. The variables from the $\sum\limits_{i\in\Gamma}|\Act_i|$ quantifications in \cite{DBLP:journals/iandc/LaroussinieM15} are mutually exclusive within each $\Act_i$. Observe that $|\Act_i|$ propositional variables $a_{i,1},\ldots,a_{i,|\Act_i|}$ with the condition that every state satisfies exactly one of them can be expressed as elementary conjunctions in terms of $m_i\defeq\lceil\log_2|\Act_i|\rceil$ {\em independent} propositional variables $r_{i,1},\ldots,r_{i,m_i}$:
\[\begin{array}{lll}
a_{i,1}&\defeq&\neg r_{i,1}\wedge\neg r_{i,2}\wedge\ldots\wedge\neg r_{i,m_i}\\
a_{i,2}&\defeq&r_{i,1}\wedge\neg r_{i,2}\wedge\ldots\wedge\neg r_{i,m_i}\\
a_{i,3}&\defeq&\neg r_{i,1}\wedge r_{i,2}\wedge\ldots\wedge\neg r_{i,m_i}\\
a_{i,4}&\defeq& r_{i,1}\wedge r_{i,2}\wedge\ldots\wedge\neg r_{i,m_i}\\
\ldots
\end{array}
\]
Let $\bigvee\limits_{k=1}^{|\Act_i|}a_{i,k}$ be written as a Boolean combination of $r_{i,1},\ldots,r_{i,m_i}$. Then
$\exists a_{i,1}\ldots\exists a_{i,|\Act_i|} A$ can be written as\footnote{Here and in the sequel $[X/Y]F$ stands for the result of replacing $Y$ by $X$ in $F$. $[X_1/Y_1,X_2/Y_2,\ldots]F$ and $[X_k/Y_k:k\in\ldots]F$ similarly denote multiple simultaneous replacements.}
\[\exists r_{i,1}\ldots\exists r_{i,m_i}\bigl(\bigvee\limits_{k=1}^{|\Act_i|}a_{i,k}\wedge[\neg r_{i,1}\wedge\neg r_{i,2}\wedge\ldots\wedge\neg r_{i,m_i}/a_{i,1},\ r_{i,1}\wedge\neg r_{i,2}\wedge\ldots\wedge\neg r_{i,m_i}/a_{i,2},\ \ldots\ ]A\bigr).\]
This means that the translation from \cite{DBLP:journals/iandc/LaroussinieM15} can be optimized to require $\sum\limits_{i\in\Gamma}\lceil\log_2|\Act_i|\rceil$ quantifications for encoding actions, which equals $|\Gamma|$ only if the players have exactly two actions each. The reduced number of quantifications comes at the cost of using the elementary conjuctions in terms of $r_{i,1},\ldots,r_{i,m_i}$ wherever a single $a_i$ would do otherwise. However, this is not always a $\lceil\log_2|\Act_i|\rceil$-blowup. For instance, $\bigvee\limits_{k=1}^{|\Act_i|}a_k$ can be written in terms of $r_{i,1},\ldots,r_{i,m_i}$ as a formula whose length is linear in $\lceil\log_2|\Act_i|\rceil$ and not in $|\Act_i|\lceil\log_2|\Act_i|\rceil$. See Section \ref{complexity} on using the same optimization for finitely encoding updatable preference.

The transition from mutually exclusive to independent variables is not new. It can be regarded as taking place `automagically', if model checking is done symbolically: the subsets of an $N$-element finite domain take $\lceil\log_2 N\rceil$ variables represent as BDDs. We dwell on the possibility to move from mutually exclusive to independent variables because this gives us better estimates on the complexity of translations that are natural to invent in terms of mutually exclusive variables. There is more on this in Section \ref{complexity}.

\subsection{Further Reduction of the Number of Quantifications}

As mentioned in the beginning of this section, translating a formula of the form $\atlDclm{\Gamma}B$ into $\QCTL^*$ in the proposed way takes $|\Gamma|$ quantified propositional variables. Let $B$ have no subformulas of the form $\atlDclm{\Gamma'}B'$ such that both $\Gamma'\cap\Delta\not=\emptyset$ and $\Gamma'\setminus\Delta\not=\emptyset$, that is, assume that the members of $\Gamma$ are never treated separately in what $B$ says. Then $\Gamma$ can be regarded as a single player within $B$, with the combined powers of its members and its actions being $\angles{a_i:i\in\Gamma}\in\prod\limits_{i\in\Gamma}\Act_i$. For such $B$, letting $\{i_k,\ldots,i_m\}\defeq\Gamma$, a clause about $t_{i_1,\ldots,i_n}^{p_1,\ldots,p_n}(\atlDclm{\Gamma}B)$ that introduces just {\em one} quantified variable $q_\Gamma$ for the whole of $\Gamma$ can be written as follows:
\[
t_{i_1,\ldots,i_n}^{p_1,\ldots,p_n}(\atlDclm{\Gamma}B)\ \defeq\ 
\exists q_k\ldots\exists q_{k-1},q_\Gamma
\left(\begin{array}{l}
\forall\Box\bigg(\bigvee\limits_{\angles{a_i:i\in\Gamma}\in\Act_\Gamma}\forall\Next(q_\Gamma\Leftrightarrow\bigwedge\limits_{i\in\Gamma} a_i)\bigg)\wedge\\
\forall\left(\begin{array}{l}\Next\Box\bigg(\bigwedge\limits_{l=1}^{k-1} p_i\wedge q_\Gamma\bigg)\Rightarrow t_{i_1,\ldots,i_{k-1},\Gamma}^{p_1,\ldots,p_{k-1},q_\Gamma}(B)\end{array}\right)
\end{array}\right)\]
Observe that the translation of $B$ in this clause is written as $t_{i_1,\ldots,i_{k-1},\Gamma}^{p_1,\ldots,p_{k-1},q_\Gamma}(B)$, indicating that $\Gamma$ corresponds to the single variable $q_\Gamma$. This disables any further use of both the general clause and the optimized one on subformulas $\atlDclm{\Gamma'}B'$ of $B$ such that $\Gamma'\cap\Delta\not=\emptyset$ and $\Gamma'\setminus\Delta\not=\emptyset$.

\section{Binary Preference in $\QCTL^*$}
\label{binPrefsection}

In this section we introduce the extension $\QCTL^*_<$ of $\QCTL^*$ by a binary temporal preference modality $<$, which is a special case of global binary preference operator $<_{\forall\forall}$ on abstract possible worlds from 
\cite{DBLP:journals/fuin/HarrensteinHMW03,DBLP:journals/jphil/BenthemGR09}. In these works binary preference connectives are defined starting from a basic unary modality for preference. The use of preference as the accessibility relation gives access only to the worlds which are preferred over the reference one. Access to {\em all} worlds is provided by adding the {\em universal modality} \cite{GorankoPassy1992}. In our setting, given a reference state, the possible worlds are the outgoing infinite paths, and the universal modality is $\CTL$'s $\forall\Next$. The syntax of $\QCTL^*_<$ includes state formulas of the form $B_1<B_2$ where $B_1$ and $B_2$ are path formulas. Informally, $B_1<B_2$ holds, if all the infinite $B_2$-paths that start at the reference state are preferred over all the $B_1$-ones. To keep the use of parentheses down, $<$ is assumed to bind stronger than the other connectives. We take in account that preference is {\em stable}: it becomes updated along plays according to the progress players make in pursuit of their objectives but does not change otherwise. This evolution of preference is specific to the temporal setting. 
\begin{example} 
Let plays satisfying $\Next(\neg p\wedge \Next p)$ be initially preferred over plays that satisfy $\Next\Next(\neg p\wedge \Next p)$. Then moving to a $\neg p$-state would make play suffixes which satisfy $\Next p$ preferable over $\Next(\neg p\wedge \Next p)$-suffixes:
\[\Next\Next(\neg p\wedge \Next p)<\Next(\neg p\wedge \Next p)\Rightarrow\forall\Next(\neg p\Rightarrow \Next(\neg p\wedge \Next p)< \Next p).\]
A transition to a $p$-state would rule a play satisfying $\Next(\neg p\wedge \Next p)$ and it would remain to achieve a suffix satisfying $\Next(\neg p\wedge \Next p)$, in order to achieve an initial play satisfying $\Next\Next(\neg p\wedge \Next p)$:
\[\Next\Next(\neg p\wedge \Next p)<\Next(\neg p\wedge \Next p)\Rightarrow\forall\Next(p\Rightarrow \Next(\neg p\wedge \Next p)< \bot).\]
In both transitions `progress' is made on achieving a play that satisfies $\Next\Next(\neg p\wedge \Next p)$. In the first case the opportunity for a play satisfying $\Next(\neg p\wedge \Next p)$, which is preferable, is still there too. In the second case that opportunity becomes dashed; hence $\Next\Next(\neg p\wedge \Next p)<\Next(\neg p\wedge \Next p)$ is updated to become  $\Next(\neg p\wedge \Next p)<\bot$.
Assume that the move to a $\neg p$-state is followed by a move to a $p$-state. Then a play satisfying $\Next p$ is achieved and the opportunity to achieve a $\Next(\neg p\wedge \Next p)$ play is lost:
\[\Next(\neg p\wedge \Next p)<\Next p\Rightarrow\forall\Next(p\Rightarrow \bot<\top).\]
No objectives satisfy $\bot$. That is why $\bot$, which, in this example, indicates forfeiting the opportunity to  achieve $\Next(\neg p\wedge \Next p)$, is vacuously both preferable and preferred over any other condition. The $\top$ on the RHS of $<$ indicates that nothing more needs to be done for achieving $\Next p$, that is, an initial play satisfying $\Next(\neg p\wedge \Next p)$.
\end{example}
The stability of preference means that preference only needs to be specified in the beginning of plays.
In this section we consider the addition of a single preference operator $<$ to $\QCTL^*$ and not $\ATLSC^*$ for the sake of simplicity. The results apply to the setting of $\ATLSC^*$ too, with a dedicated $<_i$ for every player $i$. 

\subsection{$\QCTL^*_<$}
\label{qctldefsection}

To provide a meaning for $B_1<B_2$, we use Kripke models $M\defeq\angles{W,w_I,R,<,V}$ that are extended with a binary {\em preference} relation denoted by the same symbol $<$ as the modality for it. This is a common way of extending arenas to games. See, e.g., \cite{li2025rationalcapabilityconcurrentgames} for a recent example of adding preference to CGMs in a similar way. 
The satisfaction relation in $\QCTL^*_<$ has the forms $M,\prec,w\models A$ and $M,\prec,\bm{w}\models B$ 
for state formulas $A$ and $w\in W$, and path formulas $B$ and $\bm{w}\in R^\fin_M(w_I)$, respectively. Here $\prec$ is the preference relation obtained by updating the given $<$ along the path from $w_I$ to $w$, and the path to $\bm{w}^0$, respectively. In the tree models below, every state can be reached by a unique path from $w_I$. In general, satisfaction in the same $w$ may happen to be wrt different $\prec$, if $w$ can be reached along different paths. This is the reason for mentioning the relevant $\prec$ on the LHS of $\models$. In \cite{li2025rationalcapabilityconcurrentgames}, a separate preference relation is associated with every state, and the stability of preference is expressed by the constraint between the preference relations of states that are reachable from each other written as (\ref{prefdef}) below. This constraint is guaranteed to be satisfiable on tree models.

In $M,\prec,w\models A$, $\prec$ is supposed to be defined on the paths from
\[R^{-,\infin}_M(w)\defeq \{\bm{v}:w\cdot\bm{v}\in R^\infin_M(w)\}.\]
The superscript ${}^-$ in $R^{-,\infin}_M(w)$ indicates that the shared initial state $w$ of the paths to be compared is ignored. This state is a {\em done deed}. This design decision is compatible with the {\em strict interpretation} of the $\LTL$ connectives themselves, which is used in, e.g., \cite{GRH94}. We stop short of adopting the strict interpretation for all the temporal connectives because of the prevalence of the non-strict interpretation in the literature, especially on branching time. $R^{-,\fin}_M(w)\subseteq W^*$ is defined similarly to $R^{-,\infin}_M(w)$. 

Consider state $w$ and the finite path $\bm{v}\in R^{-,\fin}_M(w)$ leading from $w$ to some state $w'\defeq \bm{v}^{|\bm{v}|-1}$. Let preference $\prec$ apply in state $w$. Then the update of $\prec$ that would apply in state $w'$, if that state is reached along $\bm{v}$, is written $\prec_{\bm{v}}$ where, assuming $\bm{v}_1,\bm{v}_2\in R^{-,\infin}_M(w')$,
\begin{equation}\label{prefdef}
\bm{v}_1\prec_{\bm{v}} \bm{v}_2\defeq \bm{v}\cdot\bm{v}_1\prec \bm{v}\cdot\bm{v}_2\ .
\end{equation}
Since we assume preferences to be stable, all the preference relations involved in defining $\models$ for histories other than $w_I$ are obtained by updating the $<$ given as part of the model according to (\ref{prefdef}). This definition entails that, if $\bm{v_1}\cdot\bm{v_2}\in R^{-,\fin}_M(w)$, then
\begin{equation}\label{prefconcat}
\prec_{\bm{v}_1\cdot\bm{v}_2}=(\prec_{\bm{v}_1})_{\bm{v}_2}.
\end{equation}
The preference relation $<$, which is given in the model, is assumed to be defined on $R^\infin_M(w_I)$. To meet the requirement $\prec\subseteq R^{-,\infin}_M(w)\times R^{-,\infin}_M(w_I)$ on preference in state $w_I$, we use $\prec\defeq <_{w_I}$, an update of $<$ which accounts of $w_I$ being a done deed in the plays to be compared. 

In the clauses for $\models$ for the $\LTL$ connectives in $\QCTL^*_<$ below, we use (\ref{prefconcat}) for moving from reference preference relation $\prec$, which may be $<_{\bm{v}_1}$ for any $\bm{v}_1\in R^{-,\fin}_M(w_I)$, to the relevant $<_{\bm{v}}$ where $\bm{v}\defeq\bm{v}_1\cdot\bm{v}_2$ is the continuation of $\bm{v}_1$ by some $\bm{v}_2$, as we need to determine $<_\bm{v}=<_{\bm{v}_1\cdot\bm{v}_2}=(<_{\bm{v}_1})_{\bm{v}_2}$ knowing just $\prec=<_{\bm{v}_1}$ and $\bm{v}_2$.
As needed for the correct interpretation of quantification over state variables in $\QCTL^*$, satisfaction in $\QCTL^*_<$ is defined wrt the tree unfolding $M^T$ of the given Kripke model $M$. Preference $<^T$ in $M^T$ is derived from $M$'s $<$ using the isomorphism between paths in $M$ and paths in $M^T$. 
The defining clauses are as follows: 
\[\begin{array}{lcl}
M^T,\prec,w\not\models\bot \\
M^T,\prec,w\models p &\mbox{iff}& V^T(w,p) \\
M^T,\prec,w\models A_1\Rightarrow A_2 &\mbox{iff}& M^T,\prec,w\models A_2\mbox{ or } M^T,\prec,w\not\models A_1 \\
M^T,\prec,w\models\exists B &\mbox{iff}& M^T,\prec,\bm{w}\models B\mbox{ for some }\bm{w}\in R^\infin_{M^T}(w)\\
M^T,\prec,w\models\exists p A &\mbox{iff}& (M^T)_p^X,\prec,w\models A\mbox{ for some }X\subseteq W^T\\
M^T,\prec,w\models B_1< B_2 &\mbox{iff}& \bm{w}_1\prec \bm{w}_2\mbox{ for all }\bm{w}_1,\bm{w}_2\in R^{-,\infin}_{M^T}(w)\\
& & \mbox{such that }M^T,\prec_{\bm{w}_k^0},\bm{w}_k\models B_k,\ k=1,2.
\\
M^T,\prec,w^0w^1\ldots\models A &\mbox{iff}& M^T,\prec,w^0\models A\\
M^T,\prec,\bm{w}\models B_1\Rightarrow B_2 &\mbox{iff}& M^T,\prec,\bm{w}\models B_2\mbox{ or } M^T,\prec,\bm{w}\not\models B_1 \\
M^T,\prec,w^0w^1\ldots\models \Next B &\mbox{iff}& M^T,\prec_{w^1},w^1w^2\ldots\models B\\
M^T,\prec,w^0w^1\ldots\models \until {B_1}{B_2} &\mbox{iff}&\mbox{for some }n<\omega,\ M^T,\prec_{w^1\ldots w^n},w^nw^{n+1}\ldots\models B_2\\
& & \mbox{and }M^T,\prec_{w^1\ldots w^{k}},w^kw^{k+1}\ldots\models B_1,\ k=0,\ldots,n-1.
\end{array}\]
State formula $A$ is {\em valid} in $M$, if $M^T,<^T_{w_I^T},w_I^T\models A$. $A$ is {\em valid in $\QCTL^*_<$}, if it is valid in all CGMs $M$.
\begin{example}
The informal reading of $(A<B)<C$ is that $C$ is preferable over getting to prefer $A$ over $B$ in {\em one step}. This example highlights that the path formula operands of $<$ are evaluated at the paths which start at the reference state {\em with the reference state deleted}. 
\end{example}

\begin{quote}
To avoid the perpetual use of superscript ${}^T$, in the sequel we assume that the considered $M\defeq\angles{W,w_I,R,<,V}$ are {\em tree} Kripke models. Such models are isomorphic to their tree unfoldings.
\end{quote}

\subsection{Axioms for $<$ in $\QCTL^*_<$}

Our technical considerations below do not rely on the algebraic properties that are typical of preference relations. These properties may be diverse, as mentioned above. As it becomes clear below, the relevant instances of these properties are `data' in the finite representation that we use for $<$ in our technical results. Just as an example, typical preference-specific properties of $<$, which we do not use below, include {\em transitivity} and, for strict partial orders, {\em antireflexivity} and can be expressed by axioms such as
\begin{equation}\label{prefAx1}
B_1< B_2\wedge\exists\Next B_2\wedge B_2< B_3\Rightarrow B_1< B_3
\end{equation}
\mbox{ and }
\begin{equation}\label{prefAx2}
B_1< B_2\Rightarrow\forall\Next\neg(B_1\wedge B_2).
\end{equation}
$\Next$ appears in these axioms because $<$ is compatible with the {\em strict interpretation} of temporal connectives as known in temporal logic, whereas we are using the non-strict interpretation for $\until..$ and $\exists$: the operands of $<$ are evaluated at outgoing infinite paths with their starting states being the {\em successors} of the reference state and $\exists$ is about outgoing paths that include the reference state as their shared initial state.

The informal reading of (\ref{prefAx1}) is {\em if all the $B_1$-plays are worse than any $B_2$-play, and all the $B_3$-plays are better than any $B_2$-play, and a $B_2$-play exists, then all $B_3$-plays are better than all the $B_1$-plays.} The existence of a $B_2$-play ($\exists\Next B_2$) is key for the soundness of (\ref{prefAx1}): $B_1<\bot$ and $\bot<B_3$ are vacuously true. Axiom (\ref{prefAx2}) states that no play is preferable over itself, i.e., preference is strict.

Here follow the properties of $<$ and the respective axioms which we do use. The extensionality of $<$ is expressed by the axiom
\[\begin{array}{ll}
(<_1) & \forall\Next(B_1'\Rightarrow B_2')\wedge \forall\Next(B_1''\Rightarrow B_2'')\Rightarrow (B_2'< B_2''\Rightarrow B_1'< B_1'')
\end{array}\]
Since $M^T,\prec,w\models B_1< B_2$ is defined with $\bm{w}_1$ and $\bm{w}_2$ quantified universally, $<$ additionally satisfies the axioms:
\[\begin{array}{ll}
(<_2) & (B_1'\vee B_2')<(B_1''\vee B_2'')\Leftrightarrow\bigwedge\limits_{k_1,k_2=1}^2 B_{k_1}'<B_{k_2}''\\
(<_3) & \bot<B,\ \ \ B<\bot
\end{array}\]
Axioms (\ref{prefAx1}), (\ref{prefAx2}), $(<_1)$, $(<_2)$ and $(<_3)$ appear in the axiomatisation of an epistemic logic of preferences in \cite{Synthese/NaumovO2023}.

\subsection{Finitely-defined Preference and Eliminating $<$ at a Single State}
\label{finitelyDefinedPreferenceAtASingleState}

Consider the $\QCTL^*_<$ model $M\defeq\angles{W,w_I,R,<,V}$, which we assume to be a tree model. Let $w\in W$. Paths $\bm{w}_1,\bm{w}_2\in R^{-,\infin}_M(w)$ are {\em indiscernible wrt $\prec\subseteq R^{-,\infin}_M(w)\times R^{-,\infin}_M(w) $}, written $\bm{w}_1\sim_\prec\bm{w}_2$, if 
\[\bm{w}_1\prec \bm{v}\leftrightarrow \bm{w}_2\prec\bm{v}\mbox{ and }\bm{v}\prec\bm{w}_1\leftrightarrow \bm{v}\prec\bm{w}_2\mbox{ for all }\bm{v}\in R^{-,\infin}_M(w).\]
In words, $\bm{w}_1$ and $\bm{w}_2$ are indiscernible, if they are related in the same way to all the other paths from $R^{-,\infin}_M(w)$. Applications often lead to $\prec$ such that the quotient set $R^{-,\infin}_M(w_I)/_{\sim_\prec}$ is finite. The equivalence classes from $R^{-,\infin}_M(w_I)/_{\sim_\prec}$ can be viewed as objectives and the quotient relation $\prec{\hspace{-0.05in}}/_{\sim_\prec}$, which is defined by the equivalence
\[\bm{w}_1/_{\sim_\prec} \prec{\hspace{-0.05in}}/_{\sim_\prec}\bm{w}_2/_{\sim_\prec}\defeq \bm{w}_1\prec\bm{w}_2,\]
can be viewed as preference {\em on these objectives}.

Let $\bm{w}\in R^{-,\fin}_M(w_I)$, $\prec=<_{\bm{w}}$, and $w=\bm{w}^{|\bm{w}|-1}$. Assume that, along with being finitely many, the equivalence classes in $R^{-,\infin}_M(w)/_{\sim_\prec}$ are $\LTL$-definable, that is, there exist some $\LTL$ formulas $B_1,\ldots,B_K$ such that
\begin{equation}\label{quotDef}
R^{-,\infin}_M(w)/_{\sim_\prec}=\{\sem{B_k}_{M,\prec,w}:k=1,\ldots,K\}
\end{equation}
where
\[\sem{B}_{M,\prec,w}\defeq\{\bm{w}\in R^{-,\infin}_{M}(w):M,\prec_{\bm{w}^0},\bm{w}\models B\}.\]
If no two formulas define the same class from $R^{-,\infin}_M(w)/_{\sim_\prec}$, then $B_1,\ldots,B_K$ can be chosen so that $\vdash\neg(B_{k_1}\wedge B_{k_2})$ for $k_1\not=k_2$ and $\vdash\bigvee\limits_{k=1}^K B_k$ in the logic $\LTL$, and not just $\sem{B_{k_1}}_{M,\prec,w}\cap\sem{B_{k_2}}_{M,\prec,w}=\emptyset$ and $\bigcup\limits_{k=1}^K\sem{B_k}_{M,\prec,w}=\bigcup R^{-,\infin}_{M}(w)/_{\sim_\prec}=R^{-,\infin}_{M}(w)$ in $M$. This means that $B_1,\ldots,B_K$ form a {\em full system}. Given some arbitrary path formulas $B'$ and $B''$, at $M,\prec,w$ we have
\[\begin{array}{ll}
(<_{B_1,\ldots,B_K}) &\exists\Next(B'\wedge  B_{k_1})\wedge\exists\Next(B''\wedge  B_{k_2})\Rightarrow ((B'\wedge  B_{k_1})<(B''\wedge  B_{k_2})\Leftrightarrow B_{k_1}<B_{k_2})
\end{array}\]
because $\prec$ holds for either all or none of the pairs of paths that satisfy $B_{k_1}$ and $B_{k_2}$, respectively. 

Now consider the chain of equivalences
\begin{equation}\label{chain1}
\begin{array}{lcll}
B'<B'' 
&\Leftrightarrow& 
\bigg(B'\wedge \bigvee\limits_{k_1=1}^K B_{k_1}\bigg)<\bigg(B''\wedge \bigvee\limits_{k_2=1}^K B_{k_2}\bigg) &\mbox{by }(<_1)\mbox{ and }\vdash\bigvee\limits_{k=1}^K B_k\\
&\Leftrightarrow& 
\bigg(\bigvee\limits_{k_1=1}^K (B'\wedge  B_{k_1})\bigg)<\bigg(\bigvee\limits_{k_2=1}^K (B''\wedge B_{k_2})\bigg)&\mbox{by }(<_1)\\
&\Leftrightarrow& 
\bigwedge\limits_{k_1,k_2=1}^K(B'\wedge  B_{k_1})<(B''\wedge  B_{k_2})&\mbox{by }(<_2)\\
&\Leftrightarrow& 
\bigwedge\limits_{k_1,k_2=1}^K
\left(\begin{array}{l}
\exists\Next(B'\wedge  B_{k_1})\wedge\exists\Next(B''\wedge  B_{k_2})\\
\ \ \ \Rightarrow (B'\wedge  B_{k_1})<(B''\wedge  B_{k_2})
\end{array}\right)
&\mbox{by }(<_1)\mbox{ and }(<_3)\\
&\Leftrightarrow& 
\bigwedge\limits_{k_1,k_2=1}^K
\left(\begin{array}{l}
\exists\Next(B'\wedge  B_{k_1})\wedge\exists\Next(B''\wedge  B_{k_2})\\
\ \ \ \Rightarrow( B_{k_1}< B_{k_2})
\end{array}\right)
&\mbox{by }(<_{B_1,\ldots,B_K})
\end{array}
\end{equation}
This chain shows that, if $B_1,\ldots,B_K$ as described are available at $M,\prec,w$, then the use of $<$ can be restricted to operands from among $B_1,\ldots,B_K$. Furthermore, given the set 
\[P\defeq\{\angles{k_1,k_2}\in \{1,\ldots,K\}^2:M,\prec,w\models B_{k_1}< B_{k_2}\}\]
of the pairs $k_1,k_2$ for which $M,\prec,w\models B_{k_1}< B_{k_2}$, $M,\prec,w\models B'<B''$ can be expressed with no use of $<$ at all: Let
\begin{equation}\label{prefElim}
(B'<B'')_{B_1,\ldots,B_K,P}\defeq \bigwedge\limits_{\angles{k_1,k_2}\in\{1,\ldots,K\}^2\setminus P}\neg(\exists\Next(B'\wedge B_{k_1})\wedge\exists\Next(B''\wedge B_{k_2}))\ .
\end{equation}
Then obviously $M,\prec,w\models B'<B''\Leftrightarrow (B'<B'')_{B_1,\ldots,B_K,P}$.
This means that, giving a list of $\LTL$ formulas $B_1,\ldots,B_K$ such that (\ref{quotDef}) holds for $w=w_I$, and stating whether $M,<_{w_I},w_I\models B_{k_1}<B_{k_2}$ for every pair of indices $k_1,k_2=1,\ldots,K$, can serve as the description for $<$ in $M$. In Section \ref{propagate} below we show how the availability of $B_1,\ldots,B_K$ for $<_{w_I},w_I$ propagates to all reachable states and the respective updates of $<$ by virtue of the stability of $<$.

The `data' provided by $P$ should be consistent with the algebraic properties of the adopted kind of preference such as being a partial order or a pre-order. Algebraic properties such as these may facilitate high-level reasoning using axioms like (\ref{prefAx1}) and (\ref{prefAx2}) but, as long as $R^{-,\infin}_{M}(w_I)/_{\sim_{<}}$ is finite and the equivalence classes wrt $\sim_<$ in it are $\LTL$-definable, the possibility to express $B'<B''$ as $(B'<B'')_{B_1,\ldots,B_K,P}$ does not depend on these properties.

\subsection{The Stability of $<$ and Eliminating Nested Occurrences of $<$}
\label{propagate}

Determining whether, e.g., $M,\prec,\bm{w}\models\until{F}{G}$, requires evaluating $F$ and $G$ at some sequence of suffixes of $\bm{w}$ wrt the respective updates of $\prec$. In this section and the following Section \ref{elimAll} we show how the availability of $B_1,\ldots,B_K$ and $P$, which enable the elimination of $<$ in $M,\prec,w\models B'<B''$ as described in Section \ref{finitelyDefinedPreferenceAtASingleState} above for a fixed $w$ and the respective $\prec$, can be propagated to enable eliminating $<$ in $M,\prec',w'\models B'< B''$ for $w'$ which are reachable from $w$, that is for $w'=\bm{v}^{|\bm{v}|-1}$, $\prec'=\prec_\bm{v}$, where $\bm{v}\in R^\fin_{M}(w)$. We consider the case of $\bm{v}=w'$, i.e., $w'\in R(w)$, first, that is the case of $\bm{v}$ consisting of just the state $w'$ and representing the step from $w$ to $w'$. 

Let $B_1,\ldots,B_K$ satisfy (\ref{quotDef}) wrt $\prec,w$. Let $B_1,\ldots,B_K$ be written in terms of the atomic propositions $p_1,\ldots,p_L$, and consider their GNFs (see Section \ref{LTLGNFs}):
\begin{equation}\label{gnf}
\bigvee\limits_{\varepsilon_1,\ldots,\varepsilon_L} \bigwedge\limits_{l=1}^L \varepsilon_l p_l\wedge \Next B_k^{\varepsilon_1,\ldots,\varepsilon_L},\ k=1,\ldots,K
\end{equation}
where, in the $2^L$ disjuncts above, $\varepsilon_1,\ldots,\varepsilon_L$ are optional negations. 
Let $\varepsilon_1,\ldots,\varepsilon_L$ be the unique list of optional negations such that $\bigwedge\limits_{l=1}^L \varepsilon_l p_l$
holds in the considered $w'\in R(w)$. Since $B_1,\ldots,B_K$ form a full system, $B_k^{\varepsilon_1,\ldots,\varepsilon_L}$, $k=1,\ldots,K$, form a full system too. Furthermore
\begin{equation}\label{incl}
w'\cdot\sem{B_k^{\varepsilon_1,\ldots,\varepsilon_L}}_{M,\prec',w'}\subseteq\sem{B_k}_{M,\prec,w}
\end{equation}
and, by the stability of preference, 
$\sem{B_k^{\varepsilon_1,\ldots,\varepsilon_L}}_{M,\prec',w'}$ consists of $\prec'$-indiscernible paths for every $k=1,\ldots,K$. Hence
\[R^{-,\infin}_{M}(w')/_{\sim_{\prec'}}=\{\sem{B_k^{\varepsilon_1,\ldots,\varepsilon_L}}_{M,\prec',w'}:\sem{B_k^{\varepsilon_1,\ldots,\varepsilon_L}}_{M^T,\prec',w'}\not=\emptyset,\ k=1,\ldots,K\}.\]
This means that the list of formulas 
\begin{equation}\label{updatedBs}
B_k^{\varepsilon_1,\ldots,\varepsilon_L},\ k=1,\ldots,K
\end{equation}
can be used to express $<$ wrt $\prec',w'$ as in (\ref{prefElim}). Furthermore, if we choose {\em not to remove} the formulas $B_k^{\varepsilon_1,\ldots,\varepsilon_L}$ for which $\sem{B_k^{\varepsilon_1,\ldots,\varepsilon_L}}_{M^T,\prec',w'}=\emptyset$ from (\ref{updatedBs}), then (\ref{prefElim}) can be used wrt $\prec',w'$ with (\ref{updatedBs}) in the positions of $B_1,\ldots,B_K$ and {\em the same} set $P$ of pairs of indices to indicate which formulas from (\ref{updatedBs}) are related by $<$ as for $\prec,w$.

The above argument can be cast in the form of deductions based on the axioms $(<_1)$-$(<_3)$ and two more axioms about the stability of preference over time, which is reflected in the way $\prec_{w'}$ is derived from $\prec$. The first one is
\[\begin{array}{ll}
(<_4) & (b\wedge\Next B')<(b\wedge\Next B'')\Rightarrow \forall\Next(b\Rightarrow B'<B'')
\end{array}\]
where $b$ is a modality-free formula.
The following axiom $(<_5)$ is sound only if the $\LTL$ formulas $b\wedge\Next B'$ and $b\wedge\Next B''$ in them define classes of preference-indiscernible paths.
\[\begin{array}{ll}
(<_5) & 
\exists\Next(b\wedge \exists\Next B'\wedge\exists\Next B'')\wedge\forall\Next(b\Rightarrow B'<B'')
\Rightarrow (b\wedge\Next B')<(b\wedge\Next B'')
\end{array}\]
Linear time forms of $(<_4)$ and $(<_5)$ are known from the works on unary preference modalities 
\cite{DBLP:journals/fuin/HarrensteinHMW03,DBLP:conf/kramas/DegremontK08,DBLP:journals/synthese/Kurzen09,DBLP:journals/jolli/BaskentM20} and can be related to the axioms about {\em no learning} and {\em perfect recall} in epistemic temporal logic \cite{HBE2015}. As it becomes clear below, axioms $(<_4)$ and $(<_5)$ are tailored to be used on the disjuncts of GNFs (cf. Section \ref{LTLGNFs}).

Let $\varepsilon_1,\ldots,\varepsilon_L$ be such that
$\sem{B_k^{\varepsilon_1,\ldots,\varepsilon_L}}_{M,\prec',w'}\not=\emptyset$, $k=k_1,k_2$. Let $b\defeq\bigwedge\limits_{l=1}^L \varepsilon_l p_l$ for the sake of brevity. 
For the forward implication, observe that
\begin{equation}\label{forward}
\underbrace{\exists\Next(b\wedge \exists\Next B_{k_1}^{\varepsilon_1,\ldots,\varepsilon_L}\wedge\exists\Next B_{k_2}^{\varepsilon_1,\ldots,\varepsilon_L})}_{F_0}\Rightarrow
\left(\begin{array}{l}
\forall\Next(b\Rightarrow B_{k_1}^{\varepsilon_1,\ldots,\varepsilon_L}<B_{k_2}^{\varepsilon_1,\ldots,\varepsilon_L})
\Rightarrow\\
(b\wedge\Next B_{k_1}^{\varepsilon_1,\ldots,\varepsilon_L})<(b\wedge\Next B_{k_2}^{\varepsilon_1,\ldots,\varepsilon_L}).
\end{array}\right)
\end{equation}
is the instance of $(<_5)$ for the disjuncts $B_k^{\varepsilon_1,\ldots,\varepsilon_L}$ from the GNF (\ref{gnf}) of $B_k$, $k=k_1,k_2$, as $B'$ and $B''$, respectively.
Since $b\wedge\Next B_{k}^{\varepsilon_1,\ldots,\varepsilon_L}$ is a disjunct of the GNF of $B_{k}$, we have
\[\forall\Next(b\wedge\Next B_{k}^{\varepsilon_1,\ldots,\varepsilon_L}\Leftrightarrow b\wedge\Next B_{k}^{\varepsilon_1,\ldots,\varepsilon_L}\wedge B_k),\ k=k_1,k_2.\]
Using the validity of this formula, $(<_1)$, (\ref{forward}) and the instance 
\[\begin{array}{l}
\underbrace{\exists\Next(b\wedge\Next B_{k_1}^{\varepsilon_1,\ldots,\varepsilon_L}\wedge  B_{k_1})\wedge\exists\Next(b\wedge\Next B_{k_2}^{\varepsilon_1,\ldots,\varepsilon_L}\wedge  B_{k_2})}_{F_1}\Rightarrow\\
\ \ \ \ \ \ \ \ \ \ \ \ ((b\wedge\Next B_{k_1}^{\varepsilon_1,\ldots,\varepsilon_L}\wedge  B_{k_1})<(b\wedge\Next B_{k_2}^{\varepsilon_1,\ldots,\varepsilon_L}\wedge  B_{k_2})\Leftrightarrow B_{k_1}<B_{k_2})
\end{array}
\]
of $(<_{B_1,\ldots,B_k})$ for $B'$ and $B''$ being $b\wedge\Next B_{k_1}^{\varepsilon_1,\ldots,\varepsilon_L}$ and $b\wedge\Next B_{k_2}^{\varepsilon_1,\ldots,\varepsilon_L}$, respectively, we derive
\[F_0\wedge F_1\Rightarrow \forall\Next(b\Rightarrow B_{k_1}^{\varepsilon_1,\ldots,\varepsilon_L}<B_{k_2}^{\varepsilon_1,\ldots,\varepsilon_L})\Rightarrow B_{k_1}<B_{k_2}.\]
Finally, since both $M,\prec,w\models F_0$ and $M,\prec,w\models F_1$ follow from the requirement $\sem{B_k^{\varepsilon_1,\ldots,\varepsilon_L}}_{M,\prec',w'}\not=\emptyset$, $k=k_1,k_2$,
we conclude that
\[M,\prec,w\models\forall\Next(b\Rightarrow B_{k_1}^{\varepsilon_1,\ldots,\varepsilon_L}<B_{k_2}^{\varepsilon_1,\ldots,\varepsilon_L})\Rightarrow B_{k_1}<B_{k_2}.\]
For the converse implication consider the chain of implications and equivalences
\[\begin{array}{lll}B_{k_1}<B_{k_2}
&\Rightarrow 
(B_{k_1}\wedge b\wedge\Next B_{k_1}^{\varepsilon_1,\ldots,\varepsilon_L})<
(B_{k_2}\wedge b\wedge\Next B_{k_2}^{\varepsilon_1,\ldots,\varepsilon_L}) & \mbox{by }(<_2)\\
&\Leftrightarrow
(b\wedge\Next B_{k_1}^{\varepsilon_1,\ldots,\varepsilon_L})<
(b\wedge\Next B_{k_2}^{\varepsilon_1,\ldots,\varepsilon_L}) & \mbox{by the definition of GNF and }(<_1)\\
&\Rightarrow
\forall\Next(b\Rightarrow B_{k_1}^{\varepsilon_1,\ldots,\varepsilon_L}<B_{k_2}^{\varepsilon_1,\ldots,\varepsilon_L}) & \mbox{by }(<_4)
\end{array}\]
The number of the formulas $B_k^{\varepsilon_1,\ldots,\varepsilon_L}$ which are required to define the indiscernibility classes from 
$R^{-,\infin}_{M}(w')/_{\sim_{\prec'}}$ is no bigger than the number $|R^{-,\infin}_{M}(w)/_{\sim_\prec}|$ of the defining formulas $B_k$ at $\prec,w$. A decrease can occur if some of the formulas $B_k^{\varepsilon_1,\ldots,\varepsilon_L}$ are unsatisfiable at the outgoing paths from the chosen $w'$, which is possible even if the respective $B_k$ is satisfiable at some outgoing path from $w$. However, if we keep the formulas which are no longer satisfiable in the list upon moving from $B_1,\ldots,B_K$, which apply at $\prec,w$, to $B_1^{\varepsilon_1,\ldots,\varepsilon_L},\ldots,B_K^{\varepsilon_1,\ldots,\varepsilon_L}$, which apply at $\prec',w'$, then the pairs of indices $\angles{k_1,k_2}\in P$ for which $M,\prec,w\models B_{k_1}< B_{k_2}$ holds, would apply for $M,\prec',w'\models B^{\varepsilon_1,\ldots,\varepsilon_L}_{k_1}< B^{\varepsilon_1,\ldots,\varepsilon_L}_{k_2}$ as well.
If one or both of $\sem{B^{\varepsilon_1,\ldots,\varepsilon_L}_{k_1}}_{M,\prec',w'}$ and $\sem{B^{\varepsilon_1,\ldots,\varepsilon_L}_{k_2}}_{M,\prec',w'}$ are empty, then $M,\prec',w'\models B^{\varepsilon_1,\ldots,\varepsilon_L}_{k_1}< B^{\varepsilon_1,\ldots,\varepsilon_L}_{k_2}$ would hold vacuously even in case $\angles{k_1,k_2}$ is not on our list.
Hence putting
\[P'\defeq\{\angles{k_1,k_2}\in \{1,\ldots,K\}^2:M,\prec',w'\models B^{\varepsilon_1,\ldots,\varepsilon_L}_{k_1}< B^{\varepsilon_1,\ldots,\varepsilon_L}_{k_2}\}\]
implies $P'\supseteq P$, and $\angles{k_1,k_2}\in P'\setminus P$ means that at least one of $\sem{B^{\varepsilon_1,\ldots,\varepsilon_L}_{k_1}}_{M,\prec',w'}$ and $\sem{B^{\varepsilon_1,\ldots,\varepsilon_L}_{k_2}}_{M,\prec',w'}$ is $\emptyset$. According to (\ref{prefElim}), this means that, for any pair of path formulas $B'$ and $B''$, the $<$-free expression $(B'<B'')_{B^{\varepsilon_1,\ldots,\varepsilon_L}_1,\ldots,B^{\varepsilon_1,\ldots,\varepsilon_L}_K,P'}$ for $B'<B''$ in $\prec',w'$ is equivalent to $(B'<B'')_{B^{\varepsilon_1,\ldots,\varepsilon_L}_1,\ldots,B^{\varepsilon_1,\ldots,\varepsilon_L}_K,P}$, which involves $P$ instead of $P'$.

\subsection{Eliminating All Occurrences of $<$}
\label{elimAll}

Now let $w=w_I$, $\prec=<_{w_I}$ and $w'=\bm{v}^{|\bm{v}|-1}=v^N$ and $\prec'=\prec_\bm{v}$, where $\bm{v}=v^1\ldots v^N\in R^{-,\fin}_{M}(w_I)$. Let the $\LTL$ formulas $B_{I,1},\ldots,B_{I,K}$ form a full system. Let
\[R^{-,\infin}_M(w_I)/_{\sim_\prec}=\{\sem{B_{I,k}}_{M,\prec,w}:k=1,\ldots,K\}\]
and
\[P_I\defeq\{\angles{k_1,k_2}\in\{1,\ldots,K\}^2:M,<,w_I\models B_{k_1}< B_{k_2}\}.\]
Then evaluating $<$ in $v^n$ would require using $\prec_n$ where $\prec_0\defeq\prec$ and $\prec_n\defeq(\prec_{n-1})_{v^n}$, $n=1,\ldots,N$, and (\ref{prefElim}) can be used to express $M,\prec_n,v^n\models B'< B''$ using the formulas $B_{n,1},\ldots,B_{n,K}$ which are defined inductively as follows:

(i) $B_{0,k}\defeq B_{I,k}$.

\noindent
For $n>0$, let
\[\bigvee\limits_{\varepsilon_1,\ldots,\varepsilon_L} \bigwedge\limits_{l=1}^L \varepsilon_l p_l\wedge \Next B_{n-1,k}^{\varepsilon_1,\ldots,\varepsilon_L}\]
be a GNF of $B_{n-1,k}$ and $\varepsilon_1,\ldots,\varepsilon_L$ be the unique list of optional negations for which $\bigwedge\limits_{l=1}^L \varepsilon_l p_l$ holds at $v^n$. Then

(ii) $B_{n,k}\defeq  B_{n-1,k}^{\varepsilon_1,\ldots,\varepsilon_L}$.

Since $B_{n,k}\in\Cl(B_{I,k})$ for all $n$, and $\Cl(B_{I,k})$ is finite (cf. Section \ref{LTLGNFs}), only finitely many lists of formulas of the form $B_{n,1},\ldots,B_{n,K}$ are needed to enable the use of (\ref{prefElim}) in the entire $M$. As argued in Section \ref{propagate}, the same set $P=P_I$ can be used in $(B'<B'')_{B_{n,1},\ldots,B_{n,K},P}$ in all states. To enable a satisfaction-preserving translation from $\QCTL^*_<$ to $\QCTL^*$ based on (\ref{prefElim}), it remains to encode the correspondence between reachable states and the relevant lists $B_1,\ldots,B_K$. This can be done by extending the vocabulary $AP$ of considered model $M$ by the atomic propositions $q_{B_k,k}$, $B_k\in\Cl(B_{I,k})$, $k=1,\ldots,K$ with the intended meaning of $w\models q_{B_k,k}$ being that $B_k$ is must appear in position $k$ in thet $<$-free equivalent $(B'<B'')_{B_1,\ldots,B_K,P}$ of $B'<B''$ in state $w$. 

Assuming the formulas in $\Cl(B_{I,k})$ to be pairwise non-equivalent, the intended meaning of $w\models q_{B_k,k}$ leads to a unique extension of the valuations of the considered models. The satisfaction of $B'<B''$ in $w$ depends on the respective $\prec$ too; however the relevant $\prec$ is uniquely determined because the considered model is a tree one. Let $B_k^{\varepsilon_1,\ldots,\varepsilon_L}\in\Cl(B_{I,k})$ be the tail formula from the GNF (\ref{gnf}) of $B_k$ that corresponds to the specified $\varepsilon_1,\ldots,\varepsilon_L$. Let $\{q_1,\ldots,q_Z\}\defeq\{q_{B_k,k}:B_k\in\Cl(B_{I,k}),\ k=1,\ldots,K\}$. Then the extension of the valuation of $M$ in consideration can be expressed by the formula
\begin{equation}\label{qTrans}
L(q_1,\ldots,q_Z)\defeq
\bigwedge\limits_{k=1}^K q_{B_{I,k},k}\wedge\forall\Box\bigwedge\limits_{k=1}^K\left(\begin{array}{ll}\bigwedge\limits_{B_k',B_k''\in \Cl(B_{I,k}),\ B_k'\not=B_k''}\neg(q_{B_k',k}\wedge q_{B_k'',k})\wedge\\
\ \ \bigwedge\limits_{B_k\in \Cl(B_{I,k})}q_{B_k,k}\Rightarrow\forall\Next\bigwedge\limits_{\varepsilon_1,\ldots,\varepsilon_L}\bigg(\bigwedge\limits_{l=1}^L \varepsilon_l p_l\Rightarrow q_{B_k^{\varepsilon_1,\ldots,\varepsilon_L},k}\bigg)
\end{array}\right)
\end{equation}
where, for the sake of simplicity, we assume that all the formulas from $\Cl(B_{I,k})$, $k=1,\ldots,K$ are written in terms of the atomic propositions $p_1,\ldots,p_L\in AP$, and $\varepsilon_1,\ldots,\varepsilon_L$ range over all the possible combinations of optional negations.

Now consider an arbitrary $\QCTL^*_<$ state formula $A$. Let $A'$ be obtained from $A$ by replacing every subformula occurrence $B'<B''$ by $\bigvee\limits_{B_k\in \Cl(B_{I,k}),\ k=1,\ldots,K} (q_{B_1,1}\wedge\ldots\wedge q_{B_K,K}\wedge (B'<B'')_{B_1,\ldots,B_K,P_I})$ bottom up. Then we have
\begin{equation}\label{qTrans0Fin}
M,<_{w_I},w_I\models A\Leftrightarrow\exists q_1\ldots\exists q_Z(L(q_1,\ldots,q_Z)\wedge A').
\end{equation}
The formula $\exists q_1\ldots\exists q_Z(L(q_1,\ldots,q_Z)\wedge A')$ is $<$-free, that is, in $\QCTL^*$.

\subsection{On the Complexity of Eliminating $<$ for Model Checking}
\label{complexity}

\paragraph{Specifying the relevant $B_1,\ldots,B_K$ by independent variables}

Encoding the fact that $B_1,\ldots,B_K$ are the right formulas to use in (\ref{prefElim}) in $\prec,w$ in the above way takes $\sum\limits_{k=1}^K|\Cl(B_{I,k})|$ additional atomic propositions. However, much like the variables used to denote actions in the translation from $\ATLSC^*$ to $\QCTL^*$ given in \cite{DBLP:conf/concur/LopesLM12,DBLP:journals/iandc/LaroussinieM15} and our translation given in Section \ref{atlscToqctl} of this paper, for every $k=1,\ldots,K$, every reachable state satisfies exactly one of the variables $q_{B_k,k}$. This means that, for every $k$, $q_{B_k,k}$, $B_k\in\Cl(B_{I,k})$ can be replaced by $\lceil\log_2|\Cl(B_{I,k})|\rceil$ many {\em independent} variables. 
Let $\{B_{k,1},\ldots,B_{k,|\Cl(B_{I,k})|}\}\defeq\Cl(B_{I,k})$ and consider substituting $q_{B_k,k}$ variables by distinct elementary conjunctions of some fresh variables $r_{1,k},\ldots,r_{\lceil\log_2|\Cl(B_{I,k})|\rceil,k}$ in (\ref{qTrans}). 

Doing so renders it unnecessary to have an expression such as
\[\bigwedge\limits_{B_k',B_k''\in \Cl(B_{I,k}),\ B_k'\not=B_k''}\neg(q_{B_k',k}\wedge q_{B_k'',k})\]
in (\ref{qTrans}) for specifying that no two $q_{B_k,k}$ hold in the same state. 
Now, assuming that $\{r_1,\ldots,r_Z\}\defeq\{r_{s,k}:s=1,\ldots,\lceil\log_2|\Cl(B_{I,k})|\rceil,k=1,\ldots,K\}$, where, consequently, $Z=\sum\limits_{k=1}^K\lceil\log_2|\Cl(B_{I,k})|\rceil$, (\ref{qTrans}) can be written as
\[
L(r_1,\ldots,r_Z)\defeq
\bigwedge\limits_{k=1}^K q_{B_{I,k},k}\wedge\forall\Box\bigwedge\limits_{k=1}^K\left(
\ \ \bigwedge\limits_{B_k\in \Cl(B_{I,k})}q_{B_k,k}\Rightarrow\forall\Next\bigwedge\limits_{\varepsilon_1,\ldots,\varepsilon_L}\bigg(\bigwedge\limits_{l=1}^L \varepsilon_l p_l\Rightarrow q_{B_k^{\varepsilon_1,\ldots,\varepsilon_L},k}\bigg)\right)
\]
yet with $q_{B_k,k}$ in it standing for their corresponding elementary conjunctions in terms of $r_{1,k},\ldots,r_{\lceil\log_2|\Cl(B_{I,k})|\rceil,k}$.
To write a variant of the $\QCTL^*$ translation $\exists q_1\ldots\exists q_Z(L(q_1,\ldots,q_Z)\wedge A')$ of a given $\QCTL^*_<$ formula $A$ in terms of $r_{k,s}$, we must take in account that, unless $|\Cl(B_{I,k})|$ is a power of $2$, replacing
\[\exists q_{B_{k,1},k}\ldots \exists q_{B_{k,|\Cl(B_{I,k})|},k}\mbox{ by }
\exists r_{1,k}\ldots\exists r_{\lceil\log_2|\Cl(B_{I,k})|\rceil,k}\]
would allow combinations of truth values for $r_{1,k},\ldots,r_{\lceil\log_2|\Cl(B_{I,k})|\rceil,k}$ which do not correspond to a $q_{B_k,k}$. Therefore $\exists r_{1,k}\ldots\exists r_{\lceil\log_2|\Cl(B_{I,k})|\rceil,k}$ should be used together with a restricting expression for $\bigvee\limits_{B_k\in\Cl(B_{I,k})}q_{B_k,k}$ in terms of $r_{1,k},\ldots,r_{\lceil\log_2|\Cl(B_{I,k})|\rceil,k}$. This expression takes a formula whose length is linear in $\lceil\log_2|\Cl(B_{I,k})|\rceil$ and is needed only in the initial state. $L(r_1,\ldots,r_Z)$ implies its propagation. Hence the $\QCTL^*$ translation of $\QCTL^*_<$ formula $A$ can be written as
\[
\left(\begin{array}{l}
\exists r_{1,1}\ldots\exists r_{\lceil\log_2|\Cl(B_{I,1})|\rceil,1}\\
\ldots\\
\exists r_{1,K}\ldots\exists r_{\lceil\log_2|\Cl(B_{I,K})|\rceil,K}
\end{array}\right)\left(\bigwedge\limits_{k=1}^K\left(\bigvee\limits_{B_k\in\Cl(B_{I,k})}q_{B_k,k}\right)\wedge L(r_1,\ldots,r_Z)\wedge A'\right)
\]
where $A'$ is obtained from $A$ like previously, except that all the designated occurrences of $q_{B_k,k}$ stand for their corresponding elementary conjunctions in terms of $r_{k,1},\ldots,r_{\lceil\log_2|\Cl(B_{I,k})|\rceil,k}$.
 
\paragraph{Avoiding propositional quantification for model checking}

Along with preserving satisfaction at particular models, the translation from $\QCTL^*_<$ to $\QCTL^*$ preserves validity on classes of models $\angles{W,w_I,R,<,V}$ that admit a finite description of their preference relations $<$ in terms of the same fixed list of $\LTL$ formulas $B_{I,1},\ldots,B_{I,K}$. The use of $L(q_1,\ldots,q_Z)$ as defined in (\ref{qTrans}) appears to be indispensable for validity preservation. However, this is not the case with model checking. Given a model $M$, (\ref{qTrans}) defines an assignment of the propositional variables $q_1,\ldots,q_Z$ involved in eliminating the preference operator that is unique on the reachable states of the model. This means that, model checking $\exists q_1\ldots\exists q_Z(L(q_1,\ldots,q_Z)\wedge A')$ in the given $M$ can be replaced by model checking $A'$ in an $M'\defeq \angles{W',w_I',R',V'}$ which is appropriately extended to accommodate the assignment of $q_1,\ldots,q_Z$ determined by $L(q_1,\ldots,q_Z)$. Along with rendering the use of $L(q_1,\ldots,q_Z)$ unnecessary, this implies that, formulas $\QCTL^*_<$, formulas $A$ with no propositional quantification in them can be translated into $\QCTL^*$ without the use of propositional quantification, i.e., model checking $\CTL^*_<$ reduces to model checking $\CTL^*$ without the `Q'. The details below are based on the translation from $\QCTL^*_<$ to $\QCTL^*$ which involves the variables $q_{B_k,k}$, $B_k\in\Cl(B_{I,k})$, $k=1,\ldots,K$, for the sake of simplicity. The variant of this translation based on the variables $r_{k,s}$, $s=1,\ldots,\lceil\log_2|\Cl(B_{I,k})|\rceil$, $k=1,\ldots,K$, described in this Section can be used just as well.

Given $M\defeq\angles{W,w_I,R,<,V}$ and $\bm{B}\defeq\angles{B_{I,1},\ldots,B_{I,K},P_I}$ to describe $<$ as above, we define the Kripke model $M_{\bm{B}}\defeq\angles{W_{\bm{B}},w_{{\bm{B}},I},R_{\bm{B}},V_{\bm{B}}}$, which has $AP\,\uplus\,\{q_1,\ldots,q_Z\}$ as its vocabulary so that $M_{\bm{B}},w_I\models L(q_1,\ldots,q_Z)$. Observe that $M_{\bm{B}}$ is meant for the translations of $\QCTL^*_<$ formulas into $\QCTL^*$ and therefore has no preference relation. The definition follows what is stated by $L(q_1,\ldots,q_Z)$ in the expected way:
\[
W_{\bm{B}} \ \defeq\  W\times {\mathcal P}(\{q_1,\ldots,q_Z\})\mbox{ and }
w_{{\bm{B}},I}\ \defeq\  \angles{w_I,\{q_{B_{I,k},k}:k=1,\ldots,K\}}
\]
Let $Q'\defeq\{q_{B_k,k}:k=1,\ldots,K\}$ and $B_k\in\Cl(B_{I,k})$ have the GNFs
\[\bigvee\limits_{\varepsilon_1,\ldots,\varepsilon_L}\bigwedge\limits_{l=1}^L\varepsilon_l p_l\wedge \Next B_k^{\varepsilon_1,\ldots,\varepsilon_L}\] 
where $p_1,\ldots,p_L$ are all the atomic propositions occurring in $B_{I,1},\ldots,B_{I,K}$, and $B_k^{\varepsilon_1,\ldots,\varepsilon_L}\in\Cl(B_{I,k})$, $k=1,\ldots,K$. Let $w''\in W$ and $\varepsilon_1,\ldots,\varepsilon_L$ be the unique list of optional negations satisfying $w''\models\bigwedge\limits_{l=1}^L\varepsilon_l p_l$. Then
\[R_{\bm{B}}(\angles{w',Q'},\angles{w'',Q''})\ \defeq\ R(w',w'')\wedge Q''=\{q_{B_k^{\varepsilon_1,\ldots,\varepsilon_L},k}:k=1,\ldots,K\}.\]
Finally,
\[
V_{\bm{B}}(\angles{w, Q},p)\ \defeq\ 
\left\{\begin{array}{lp{3in}}
V(w,p), & if $p\in AP$;\\
p\in Q, & if $p\in\{q_1,\ldots,q_Z\}$.
\end{array}\right.
\]

\subsection{Other Binary Preference Modalities}
The preference modality $<$ of $\QCTL^*_<$ is a temporal form of the $<_{\forall\forall}$ from \cite{DBLP:journals/jphil/BenthemGR09}. Other strict variants of the modality have been defined in \cite{DBLP:journals/jphil/BenthemGR09} by changing one or both of the universal quantifications in the definitions to existential ones. Here follow the definitions of the $\QCTL^*_<$ variants of these other preference operators: 
\[\begin{array}{lcl}
M,\prec,w\models B_1<_{\exists\forall} B_2 &\mbox{iff}& (\exists\bm{w}_1\in \sem{B_1}_{M,\prec,w})(\forall\bm{w}_2\in \sem{B_2}_{M,\prec,w})(\bm{w}_1\prec\bm{w}_2)\\
M,\prec,w\models B_1<_{\forall\exists} B_2 &\mbox{iff}& (\forall\bm{w}_1\in \sem{B_1}_{M,\prec,w})(\exists\bm{w}_2\in \sem{B_2}_{M,\prec,w})(\bm{w}_1\prec\bm{w}_2)\\
M,\prec,w\models B_1<_{\exists\exists} B_2 &\mbox{iff}& (\exists\bm{w}_1\in \sem{B_1}_{M,\prec,w})(\exists\bm{w}_2\in \sem{B_2}_{M,\prec,w})(\bm{w}_1\prec\bm{w}_2)\\
M,\prec,w\models B_1>_{\exists\forall} B_2 &\mbox{iff}& (\exists\bm{w}_1\in \sem{B_1}_{M,\prec,w})(\forall\bm{w}_2\in \sem{B_2}_{M,\prec,w})(\bm{w}_2\prec\bm{w}_1)\\
M,\prec,w\models B_1>_{\forall\exists} B_2 &\mbox{iff}& (\forall\bm{w}_1\in \sem{B_1}_{M,\prec,w})(\exists\bm{w}_2\in \sem{B_2}_{M,\prec,w})(\bm{w}_2\prec\bm{w}_1)
\end{array}\]
Some of these modalities admit axioms like $(<_2)$ for disjunction occurring in just one of the operands:
\[\begin{array}{ll}
(<_{2,\exists\forall}) & (B_1'\vee B_2')\bowtie B''\Leftrightarrow\bigvee\limits_{k_1=1}^2 B_{k_1}'\bowtie B''\mbox{ for }\bowtie=<_{\exists\forall},>_{\exists\forall}\\
(<_{2,\forall\exists}) & (B_1'\vee B_2')\bowtie(B_1''\vee B_2'')\Leftrightarrow\bigwedge\limits_{k_1=1}^2\bigvee\limits_{k_2=1}^2 B_{k_1}'\bowtie B_{k_2}''\mbox{ for }\bowtie=<_{\forall\exists},>_{\forall\exists}\\
(<_{2,\exists\exists}) & (B_1'\vee B_2')<_{\exists\exists}(B_1''\vee B_2'')\Leftrightarrow\bigvee\limits_{k_1=1}^2\bigvee\limits_{k_2=1}^2 B_{k_1}'<_{\exists\exists}B_{k_2}''\\
\end{array}\]
However, if $B_1,\ldots,B_K$ as above are available, then all the variants of binary preference can be defined in terms of $B_{k_1}<_{\forall\forall} B_{k_2}$, $k_1,k_2\in\{1,\ldots,K\}$, using the validity of the equivalences:
\[\begin{array}{l}
B'<_{\exists\forall} B''\Leftrightarrow\bigvee\limits_{k_1=1}^K\bigg(\exists\Next(B'\wedge B_{k_1})\wedge\bigwedge\limits_{k_2=1}^K(\exists\Next(B''\wedge B_{k_2})\Rightarrow B_{k_1}<_{\forall\forall} B_{k_2})\bigg)\\
B'<_{\forall\exists} B''\Leftrightarrow\bigwedge\limits_{k_1=1}^K\bigg(\exists\Next(B'\wedge B_{k_1})\Rightarrow\bigvee\limits_{k_2=1}^K(\exists\Next(B''\wedge B_{k_2})\wedge B_{k_1}<_{\forall\forall} B_{k_2})\bigg)\\
B'<_{\exists\exists} B''\Leftrightarrow\bigvee\limits_{k_1,k_2=1}^K(\exists\Next(B'\wedge B_{k_1})\wedge\exists\Next(B''\wedge B_{k_2})\wedge B_{k_1}<_{\forall\forall} B_{k_2})\\
B'>_{\exists\forall} B''\Leftrightarrow\bigvee\limits_{k_1=1}^K\bigg(\exists\Next(B'\wedge B_{k_1})\wedge\bigwedge\limits_{k_2=1}^K(\exists\Next(B''\wedge B_{k_2})\Rightarrow B_{k_2}<_{\forall\forall} B_{k_1})\bigg)\\
B'>_{\forall\exists} B''\Leftrightarrow\bigwedge\limits_{k_1=1}^K\bigg(\exists\Next(B'\wedge B_{k_1})\Rightarrow\bigvee\limits_{k_2=1}^K(\exists\Next(B''\wedge B_{k_2})\wedge B_{k_2}<_{\forall\forall} B_{k_1})\bigg)
\end{array}\]
This is so because the lack of distributivity wrt the second operand, which makes $(<_{2,\exists\forall})$ insufficient to use in a chain of equivalences like (\ref{chain1}), can be compensated for by using the validity of the equivalences
\[\begin{array}{l}
(B'\wedge B_k)<_{\exists\forall} B''\Leftrightarrow\exists\Next(B'\wedge B_k)\wedge (B'\wedge B_k)<_{\forall\forall} B''\\
(B'\wedge B_k)>_{\exists\forall} B''\Leftrightarrow\exists\Next(B'\wedge B_k)\wedge B''<_{\forall\forall}(B'\wedge B_k)
\end{array}\]
and the distributivity of $<_{\forall\forall}$.
These equivalences are valid because the paths which satisfy a $B_k$ are indiscernible wrt preference. Hence having $B_1,\ldots,B_K$ allows the expressiveness of all the variants of binary preference above to be achieved using just the operator $<=<_{\forall\forall}$ of $\QCTL^*_<$. 

\section{Second Order Quantification over Paths and Preference}
\label{pathQuantification}


$\CTL$'s path quantifier $\exists$ can be viewed as an (implicit) first-order quantifier on outgoing paths as it is about the existence of individual paths satisfying its operand formula. Unlike that, in this section we propose a specific preference related form of quantification over {\em sets} of paths. This means introducing bound symbols with the status of path formulas. E.g., we want to be able to write
\begin{equation}\label{exquant}
M,\prec,w\models\existsSim\mathbf{p}(\exists\Next \mathbf{p}\wedge B'<\mathbf{p}\wedge \mathbf{p}< B''),
\end{equation}
which states that there exist some paths starting from $w$ (with $w$ itself deleted) which are better than any $B'$-path and worse than any $B''$-path. Properties like the above can be expressed with $\existsSim$ causing $\mathbf{p}$ to range over sets of paths which are {\em closed under preference indiscernibility}. This restriction comes at hardly any technical cost: in this section we show how the elimination technique for $<$ works for $\existsSim\mathbf{p}$ too. On the other hand, if $\existsSim\mathbf{p}A$ has no occurrences of $\mathbf{p}$ but in operands of $<$, possibly in the scope of $\wedge$ and $\vee$, then letting $\mathbf{p}$ range over all sets of paths is equivalent to letting $\mathbf{p}$ range only over sets of paths that are closed under preference indiscernibility. Furthermore, it is natural to assume that any player's objectives are closed under preference indiscernibility wrt that player's preference relation. Hence the proposed form of quantification can facilitate expressing a variety of preference-related properties.

Before presenting the semantics of $\existsSim$ formally, we give an example of how it can be eliminated.
Consider the example formula (\ref{exquant}) and let $B_1,\ldots,B_K$ define the classes of $\prec$-indiscernible paths in $R^{-,\infin}_M(w)$ where $w\in W$ and $\prec$ is the preference relation corresponding to $w$ in $M\defeq\angles{W,w_I,R,<,V}$. Then, in $\prec,w$, (\ref{exquant}) is equivalent to
\[\bigvee\limits_{X\subseteq\{1,\ldots,K\}}\exists\Next\bigg(\bigvee\limits_{x\in X}B_x\bigg)\wedge B'<\bigg(\bigvee\limits_{x\in X}B_x\bigg)\wedge \bigg(\bigvee\limits_{x\in X}B_x\bigg)<B''.\]
To realize this, observe how this formula has been obtained from (\ref{exquant}): $\existsSim$ of $\existsSim\mathbf{p}$ has been replaced by a parameterized disjunction, with the parameter $X$ ranging over the subsets of $\{1,\ldots,K\}$. For every value of $X$, the disjunction $\bigvee\limits_{x\in X}B_x$ is a path formula which defines the corresponding union $\bigcup\limits_{x\in X}\sem{B_x}_{M,<,w_I}$ of the preference-indiscernibility equivalence classes of paths. Every set of paths that is closed under preference indiscernibility admits this representation. Hence $\bigvee\limits_{x\in X}B_x$ define the exactly the sets of paths where $\mathbf{p}$ is supposed to hold.
Preference can be eliminated from the above formula in the way described in Section \ref{binPrefsection} by further transforming it into
\[\bigvee\limits_{X\subseteq\{1,\ldots,K\}}\exists\Next\bigg(\bigvee\limits_{x\in X}B_x\bigg)\wedge\bigwedge_{k_1,k_2=1}^K\bigg(\exists\Next(B'\wedge B_{k_1})\wedge\exists\Next(B''\wedge B_{k_2})\Rightarrow B_{k_1}<\bigg(\bigvee\limits_{x\in X}B_x\bigg)\wedge \bigg(\bigvee\limits_{x\in X}B_x\bigg)<B_{k_2}\bigg),\]
where the disjunctions $\bigvee\limits_{x\in X}B_x$ are compared with whatever formulas from $B_1,\ldots,B_K$ are consistent with $B'$ and $B''$, respectively. The repeated application of $(<_2)$ allows the formula to be transformed into
\[\bigvee\limits_{X\subseteq\{1,\ldots,K\}}\exists\Next\bigg(\bigvee\limits_{x\in X}B_x\bigg)\wedge\bigwedge_{k_1,k_2=1}^K\bigwedge\limits_{x\in X}
(\exists\Next(B'\wedge B_{k_1})\wedge\exists\Next(B''\wedge B_{k_2})\Rightarrow B_{k_1}<B_x\wedge B_x<B_{k_2}),\]
where the only operands of $<$ are formulas from the list $B_1,\ldots,B_K$. Since the pairs of indices $k_1,x$ and $x,k_2$ for which $B_{k_1}<B_x$ and $B_x<B_{k_2}$ are assumed to be known and, except for some vacuous comparisons, to be the same for all $\prec,w$ the given $M$, this concludes the elimination of $\existsSim$ from (\ref{exquant}).
 
The use of $\exists\Next\bigg(\bigvee\limits_{x\in X}B_x\bigg)$ in this example indicates that we are interested only in the non-empty unions of preference-indiscernible classes of paths. We can obviously further restrict the disjunction $\bigvee\limits_{X\subseteq\{1,\ldots,K\}}$ to range only over single classes of preference-indiscernible paths to express the same property of $B'$ and $B''$:
\[\bigvee\limits_{x=1}^K\bigwedge_{k_1,k_2=1}^K
(\exists\Next(B'\wedge B_{k_1})\wedge\exists\Next(B''\wedge B_{k_2})\Rightarrow B_{k_1}<B_x\wedge B_x<B_{k_2})\]
A quantifier that makes $\mathbf{p}$ range over the unions of indiscernibility classes would obviously subsume a quantifier that causes $\mathbf{p}$ to range over individual classes. Letting $\existsSim$ and $\existsOne$, respectively, stand for the two quantifiers, $\existsOne$ can be defined in terms of $\existsSim$ by the equivalence
\[\existsOne\mathbf{p}A\Leftrightarrow\existsSim\mathbf{p}(\exists\Next\mathbf{p}\wedge\forallSim\mathbf{q}(\exists\Next(\mathbf{p}\wedge\mathbf{q})\Rightarrow\forall\Next(\mathbf{p}\Rightarrow\mathbf{q})))\]
However, the example shows that $\existsOne$ can be eliminated a lot more efficiently.

The rest of the section gives of a formal definition of $\existsSim$ and $\existsOne$ on path variables and how these quantifiers can be eliminated in $\QCTL^*_<$ models where preference admits the finite description used to eliminate $<$ in Section \ref{finitelyDefinedPreferenceAtASingleState}. This means that model checking and validity in the extension of $\QCTL^*_<$ by $\existsSim$ and $\existsOne$ can be reduced to $\QCTL^*$ just like $<$.

To provide a semantics for atomic propositions on paths, we consider Kripke models $M\defeq\angles{W,w_I,R,<,V}$ with the vocabulary $AP$ being a disjoint union $AP_{\mathit{state}}\cup AP_{\mathit{path}}$ and $V\subseteq W\times AP_{\mathit{state}}\cup\bigcup\limits_{w\in W} R_M^{-,\infin}(w)\times AP_{\mathit{path}}$. We use boldface to distinguish $\mathbf{p}\in AP_{\mathit{path}}$ and second order {\em path} quantifications $\existsSim\mathbf{p}$ from the usual kind of atomic propositions $p\in AP_{\mathit{state}}$ and {\em state} quantifications $\exists p$ that are already available in $\QCTL^*$. Given state formula $A$, $\existsSim\mathbf{p} A$ is a state formula too. The second operand of the union describing $V$ means that $V(\bm{w},\mathbf{p})$ applies to infinite paths in $M$. The defining clause for $\models$ on such variables is
\[\begin{array}{lcl}
M,\prec,\bm{w}\models \mathbf{p} &\mbox{iff}& V(\bm{w},\mathbf{p})
\end{array}\]
We define $\existsSim\mathbf{p}$ and $\existsOne\mathbf{p}$ to bind $\mathbf{p}$ to subsets of $R_M^{-,\infin}(w)$ where $w$ is the reference state, and not any subset of $\bigcup\limits_{w\in W} R_M^{-,\infin}(w)$, which includes paths with different starting states. At state $w$, $\existsOne\mathbf{p}$ causes $\mathbf{p}$ to range over the equivalence classes in $R_M^{-,\infin}(w)$ wrt preference indiscernibility, and $\existsSim\mathbf{p}$ causes $\mathbf{p}$ to range over the subsets of $R_M^{-,\infin}(w)$ which are closed under preference indiscernibility, including $\emptyset$ and the whole $R_M^{-,\infin}(w)$. Given $w\in W$, and $X\subseteq R_M^{-,\infin}(w)$, $V_{\mathbf{p}}^X$ is defined like in the case of state-based $p$, and $M_{\mathbf{p}}^X\defeq\angles{W,w_I,R,<,V_{\mathbf{p}}^X}$. The defining clauses for $\models$ on $\existsSim\mathbf{p}$- and $\existsOne\mathbf{p}$-formulas are:
\[\begin{array}{lcl}
M,\prec,w\models\existsSim \mathbf{p} A &\mbox{iff}& (M)_{\mathbf{p}}^{\cup X},\prec,w\models A\mbox{ for some }X\subseteq R_M^{-,\infin}(w)/_{\sim_\prec}\\
M,\prec,w\models\existsOne \mathbf{p} A &\mbox{iff}& (M)_{\mathbf{p}}^X,\prec,w\models A\mbox{ for some }X\in R_M^{-,\infin}(w)/_{\sim_\prec}
\end{array}\]
The paths we bind $\mathbf{p}$ to are with their shared starting state deleted. Evaluating such $\mathbf{p}$, which have the status of path formulas, is relevant only at paths which start at some successor of the state $w$ where the quantification occurs. They are trivially false elsewhere. This makes it useful to write formulas so that occurrences of $\mathbf{p}$, which are to be evaluated at paths from $R_M^{-,\infin}(w)$, appear only in the scope of a single $\Next$ and no $\until..$s. Every $\QCTL^*_<$ state formula $A$ and path formula $B$ can be translated into equivalent ones which satisfy this restriction. We denote the translation by $t_{\mathbf{p}}^1$, which indicates that it is about occurrences of $\mathbf{p}$ which are to be evaluated `{\em one} state ahead'. This translation incorporates $t_{\mathbf{p}}^0$, which is about $\mathbf{p}$ {\em at} the reference state, or, for path formulas, at the initial state of the reference path. In the clauses below, $*\in\{0,1\}$ and $\mathop{\exists}\limits_\varepsilon$ stands for either $\existsSim$ or $\existsOne$. 
\[\begin{array}{lcl}
t_{\mathbf{p}}^*(\bot) &\defeq& \bot\\
[2mm]
t_{\mathbf{p}}^*(p) &\defeq& p\\
[2mm]
t_{\mathbf{p}}^*(A_1\Rightarrow A_2) &\defeq& t_{\mathbf{p}}^*(A_1)\Rightarrow t_{\mathbf{p}}^*(A_2)\\
[2mm]
t_{\mathbf{p}}^*(\exists B) &\defeq& \exists t_{\mathbf{p}}^*(B)\\
[2mm]
t_{\mathbf{p}}^*(\exists p B) &\defeq& \exists p t_{\mathbf{p}}^*(B)\\
[2mm]
t_{\mathbf{p}}^*(\mathop{\exists}\limits_\varepsilon \mathbf{p} A) &\defeq& \mathop{\exists}\limits_\varepsilon \mathbf{p} t_{\mathbf{p}}^1(A)\\
[2mm]
t_{\mathbf{p}}^*(\mathop{\exists}\limits_\varepsilon \mathbf{q} A) &\defeq& \mathop{\exists}\limits_\varepsilon \mathbf{q}t_{\mathbf{p}}^*( t_{\mathbf{q}}^1(A)), \mbox{ if }\mathbf{q}\not=\mathbf{p}\\
[2mm]
t_{\mathbf{p}}^1(\mathbf{p}) &\defeq& \bot
\end{array} \begin{array}{lcl}
t_{\mathbf{p}}^1(\mathbf{q}) &\defeq& \mathbf{q}, \mbox{ if }\mathbf{q}\not=\mathbf{p}\\
[2mm]
t_{\mathbf{p}}^*(B_1\Rightarrow B_2) &\defeq& t_{\mathbf{p}}^*(B_1)\Rightarrow t_{\mathbf{p}}^*(B_2)\\
[2mm]
t_{\mathbf{p}}^1(\Next B) &\defeq& \Next t_{\mathbf{p}}^0(B)\\
[2mm]
t_{\mathbf{p}}^1(B_1< B_2) &\defeq& t_{\mathbf{p}}^0(B_1)<t_{\mathbf{p}}^0(B_2)\\
[2mm]
t_{\mathbf{p}}^0(\mathbf{q}) &\defeq& \mathbf{q}\mbox{ for all }\mathbf{q},\mbox{ including }\mathbf{p}\\
[2mm]
t_{\mathbf{p}}^0(\Next B) &\defeq& \Next [\bot/\mathbf{p}]B\\
[2mm]
t_{\mathbf{p}}^0(B_1< B_2) &\defeq& [\bot/\mathbf{p}]B_1<[\bot/\mathbf{p}]B_2\\
[2mm]
t_{\mathbf{p}}^*(\until{B_1}{B_2}) &\defeq& t_{\mathbf{p}}^*(B_2\vee B_1\wedge \Next \until{B_1}{B_2})\\
\end{array}\]
Here $[\bot/\mathbf{p}]$ affects only the free occurrences of $\mathbf{p}$. In words, $t_{\mathbf{p}}^0(F)$ and $t_{\mathbf{p}}^1(F)$ first transform the given formula $F$ into an equivalent one where the occurrences of $\mathbf{p}$ to be evaluated at the reference state and those to be evaluated one step ahead are separated from the other occurrences of $\mathbf{p}$. The key clause for achieving this is the one which defines the two translations on $\until..$-formulas. Then $t_{\mathbf{p}}^0(F)$ replaces all the occurrences of $\mathbf{p}$, except those to be evaluated at the reference state, by $\bot$. Similarly, $t_{\mathbf{p}}^1(F)$ replaces by $\bot$ the occurrences of $\mathbf{p}$ which are not to be evaluated at exactly one step ahead. The working of $t_{\mathbf{p}}^1(F)$ is consistent with the semantics of $\existsSim\mathbf{p}$ and $\existsOne\mathbf{p}$ because both quantifiers cause $\mathbf{p}$ to range over sets of paths which start from the reference state but have that shared initial state deleted, and, on tree models such as what we are assuming, $R_M^{-,\infin}(w')\cap R_M^{-,\infin}(w')=\emptyset$ for $w'\not=w''$. Hence $M,\prec,w\models\mathop{\exists}\limits_\varepsilon\mathbf{p} A\Leftrightarrow\mathop{\exists}\limits_\varepsilon\mathbf{p}t_{\mathbf{p}}^1(A)$ for any state formula $A$. The benefit of using $t_{\mathbf{p}}^1(.)$ is that, unlike $A$ itself in general, the occurrences of $\mathbf{p}$ in $t_{\mathbf{p}}^1(A)$, are safe to replace by formulas which define the correct subsets of $R_M^{-,\infin}(w)$. These occurrences become evaluated only at paths from $R_M^{-,\infin}(w)$, whereas the given $A$ may have $\mathbf{p}$ that can be evaluated at other infinite paths too. This enables eliminating $\existsSim\mathbf{p}$ from $\existsSim\mathbf{p}A$ by replacing the bound occurrences of $\mathbf{p}$ in $t_{\mathbf{p}}^1(A)$ by disjunctions of the formulas $B_1,\ldots,B_K$ which provide finite description of preference as proposed, and, while correctly expressing the classes of indiscernibility wrt preference at the considered state, $B_1,\ldots,B_K$ may happen to hold at paths starting from other states too, if evaluated there.

Alternatively, the effect of using $t_{\mathbf{p}}^0(F)$ and $t_{\mathbf{p}}^1(F)$ can be achieved by restricting the syntax of formulas $A$ in $\existsSim \mathbf{p}A$ in a context-sensitive way by requiring $A$ to have the syntax $A_{\mathbf{p},1}$ where $A_{\mathbf{p},k}$ restricts the occurrences of $\mathbf{p}$ in formulas to refer to some fixed number $\leq k$ of steps ahead each.
\[\begin{array}{ll}A_{\mathbf{p},0}\,::=\,\bot\,\mid\,p\,\mid\,A_{\mathbf{p},0}\Rightarrow A_{\mathbf{p},0}\,\mid\,\exists B_{\mathbf{p},0}\,\mid\,\exists p A_{\mathbf{p},0}\\  
B_{\mathbf{p},0}\,::=\,\mathbf{p}\,\mid\,A_{\mathbf{p},0}\,\mid\,B_{\mathbf{p},0}\Rightarrow B_{\mathbf{p},0}\,\mid\,\Next B_{-\mathbf{p}}\\ \\
A_{\mathbf{p},n+1}\,::=\,\bot\,\mid\,p\,\mid\,A_{\mathbf{p},n+1}\Rightarrow A_{\mathbf{p},n+1}\,\mid\,\exists B_{\mathbf{p},n+1}\,\mid\,\exists p A_{\mathbf{p},n+1}\\  
B_{\mathbf{p},n+1}\,::=\,A_{\mathbf{p},n+1}\,\mid\,B_{\mathbf{p},n+1}\Rightarrow B_{\mathbf{p},n+1}\,\mid\,\Next B_{\mathbf{p},n}\\ 
\end{array}\]
Here $A_{-\mathbf{p}}$ and $B_{-\mathbf{p}}$ stand for $\mathbf{p}$-free state and path formulas, respectively, possibly with occurrences of $\until..$. Equivalents of formulas with $\mathbf{p}$-free $\until..$-subformulas that match this syntax can be obtained by repeatedly using the validity of $\until {B'}{B''}\Leftrightarrow B''\vee B'\wedge\until{B'}{B''}$. This syntax is sufficient for $A$ with $\mathbf{p}$ as the only path variable, for the sake of simplicity. More path variables would require a separate counter $n$ each.

We incorporate the elimination of $\existsSim$ and $\existsOne$ into the translation for eliminating $<$ from Section \ref{elimAll} as follows. Recall that eliminating $<$ from formula $A$ involves replacing occurrences of $B'<B''$ in $A$ by $\bigvee\limits_{B_k\in \Cl(B_{I,k}),\ k=1,\ldots,K} (q_{B_1,1}\wedge\ldots\wedge q_{B_K,K}\wedge (B'<B'')_{B_1,\ldots,B_K,P_I})$
where $q_{B_1,1}\wedge\ldots\wedge q_{B_K,K}$ is meant to indicate that, in $\prec,w$, $R_M^{-,\infin}(w)/_{\sim_\prec}=\{\sem{B_k}_{M,\prec,w}:k=1,\ldots,K,\ \sem{B_k}_{M,\prec,w}\not=\emptyset\}$. Consider a subformula $\existsOne\mathbf{p} C$ of $A$. To eliminate $\existsOne\mathbf{p}$ from it, occurrences of $\mathbf{p}$ in $\existsOne\mathbf{p} C$ must be replaced by
\[\bigvee\limits_{B_k\in \Cl(B_{I,k}),\ k=1,\ldots,K} \bigg(q_{B_1,1}\wedge\ldots\wedge q_{B_K,K}\wedge\bigvee\limits_{k=1}^K[B_k/\mathbf{p}]t_{\mathbf{p}}^1(C)\bigg).\]
The substitution $[B_k/\mathbf{p}]$ affects only the free occurrences of $\mathbf{p}$ in its operand formula. 
The replacing formula states that $C$ must hold for $\mathbf{p}$ evaluating to one of the preference indiscernibility classes $R_M^{-,\infin}(w)/_{\sim_\prec}$, assuming that in the reference state these classes are expressed by the $\LTL$ formulas $B_1,\ldots,B_K$. Similar reasoning shows that, to eliminate $\existsSim\mathbf{p}$, $\existsSim\mathbf{p} C$ must be replaced by
\[\bigvee\limits_{B_k\in \Cl(B_{I,k}),\ k=1,\ldots,K} \bigg(q_{B_1,1}\wedge\ldots\wedge q_{B_K,K}
\wedge \bigvee\limits_{X\subseteq \{1,\ldots,K\}}\bigg[\bigg(\bigvee\limits_{k\in X}B_k\bigg)/\mathbf{p}\bigg]t_{\mathbf{p}}^1(C)\bigg).\]
In the example in Section \ref{exampleNash} we use the corresponding expression for $\forallSim\mathbf{p}C$, which stands for $\neg\existsSim\mathbf{p}\neg C$"
\begin{equation}\label{forallPathElim}
\bigwedge\limits_{B_k\in \Cl(B_{I,k}),\ k=1,\ldots,K} \bigg(q_{B_1,1}\wedge\ldots\wedge q_{B_K,K}
\Rightarrow \bigwedge\limits_{X\subseteq \{1,\ldots,K\}}\bigg[\bigg(\bigvee\limits_{k\in X}B_k\bigg)/\mathbf{p}\bigg]t_{\mathbf{p}}^1(C)\bigg).
\end{equation}
There can be nested quantifications to eliminate in the given $A$; the elimination works bottom-up. The elimination of $<$ described in Section \ref{elimAll} can be carried out either before or after the above steps to obtain the desired $<$- and $\existsSim$- and $\existsOne$-free equivalent $A'$.

\section{$\ATLSC^*$ with Preference}
\label{atlWithPreferenceSection}

For the sake of simplicity, we presented $<$ and the related form path quantification in Sections \ref{binPrefsection} and \ref{pathQuantification} in the setting of $\QCTL^*$ wrt Kripke models $M\defeq\angles{W,w_I,R,<,V}$ with just one preference relation $<$ included to provide the semantics of a single preference operator. In this section we spell out the definition of an extension $\ATLSCPref^*$ of $\ATLSC^*$ by preference operators and the related second order path quantifiers for all players. 

CGMs for $\ATLSC^*$ with preference have the form $\angles{W,w_I,\angles{\Act_i:i\in\Ag},o,\angles{<_i:i\in\Ag},V}$ where $\angles{<_i:i\in\Ag}$ are the preference relations, and all the other components are as known. Accordingly, the syntax $\ATLSCPref^*$ includes a $<_i$ and separate path quantifiers $\existsSim{}_i$ and $\existsOne{}_i$ to bind path variables which range over sets of paths that are closed under preference indiscernibility wrt $<_i$ for every $i\in\Ag$. Along with a strategy context, $\models$ requires tuple of preference relations to appear on the LHS. Here follow the defining clauses:
\[\begin{array}{lcl}
M,\rho,\angles{\prec_i:i\in\Ag},w\not\models\bot \\
M,\rho,\angles{\prec_i:i\in\Ag},w\models p &\mbox{iff}& V(w,p) \\
M,\rho,\angles{\prec_i:i\in\Ag},w\models A_1\Rightarrow A_2 &\mbox{iff}& M,\rho,\angles{\prec_i:i\in\Ag},w\models A_2\mbox{ or } M,\rho,\angles{\prec_i:i\in\Ag},w\not\models A_1 \\
M,\rho,\angles{\prec_i:i\in\Ag},w\models\atlDclm{\Gamma} B &\mbox{iff}& \mbox{there exists a strategy profile }\rho'\mbox{ for }\Gamma\mbox{ from }w,\\
&&\mbox{such that }M,\rho'\circ\rho,\angles{\prec_i:i\in\Ag},\bm{w}\models B\mbox{ for all }\bm{w}\in\out(w,\rho'\circ\rho)\\
M,\rho,\angles{\prec_i:i\in\Ag},w\models\atlCclm{\Gamma} A &\mbox{iff}& M,\rho|_{\Ag\setminus\Gamma},\angles{\prec_i:i\in\Ag},w\models A \\
M,\rho,\angles{\prec_i:i\in\Ag},w\models B_1<_i B_2 &\mbox{iff}& \bm{w}_1\prec_i \bm{w}_2\mbox{ for all }\bm{w}_1,\bm{w}_2\in R^{-,\infin}_{M}(w)\\
& & \mbox{such that }M,w^{-1}\rho,\angles{(\prec_i)_{\bm{w}_k^0}:i\in\Ag},\bm{w}_k\models B_k,\ k=1,2.\\
M,\rho,\angles{\prec_i:i\in\Ag},w\models\existsSim{}_i \mathbf{p} A &\mbox{iff}& (M)_{\mathbf{p}}^{\cup X},\prec,w\models A\mbox{ for some }X\subseteq R_M^{-,\infin}(w)/_{\sim_{\prec_i}}\\
M,\rho,\angles{\prec_i:i\in\Ag},w\models\existsOne{}_i \mathbf{p} A &\mbox{iff}& (M)_{\mathbf{p}}^X,\prec,w\models A\mbox{ for some }X\in R_M^{-,\infin}(w)/_{\sim_{\prec_i}}\\
\\
M,\rho,\angles{\prec_i:i\in\Ag},\bm{w}\models \mathbf{p} &\mbox{iff}& V(\bm{w},\mathbf{p})\\
M,\rho,\angles{\prec_i:i\in\Ag},w^0w^1\ldots\models A &\mbox{iff}& M,\rho,\angles{\prec_i:i\in\Ag},w^0\models A\\
M,\rho,\angles{\prec_i:i\in\Ag},\bm{w}\models B_1\Rightarrow B_2 &\mbox{iff}& M,\rho,\angles{\prec_i:i\in\Ag},\bm{w}\models B_2\mbox{ or } M,\rho,\angles{\prec_i:i\in\Ag},\bm{w}\not\models B_1 \\
M,\rho,\angles{\prec_i:i\in\Ag},w^0w^1\ldots\models \Next B &\mbox{iff}& M,(w^0)^{-1}\rho,\angles{(\prec_i)_{w^1}:i\in\Ag},w^1w^2\ldots\models B\\
M,\rho,\angles{\prec_i:i\in\Ag},w^0w^1\ldots\models \until {B_1}{B_2} &\mbox{iff}&\mbox{for some }n<\omega,\\
& & M,(w^0\ldots w^n)^{-1}\rho,\angles{(\prec_i)_{w^1\ldots w^n}:i\in\Ag},w^nw^{n+1}\ldots\models B_2\\
& & \mbox{and }M,(w^0\ldots w^k)^{-1}\rho,\angles{(\prec_i)_{w^1\ldots w^n}:i\in\Ag},w^kw^{k+1}\ldots\models B_1,\\
& & k=0,\ldots,n-1.
\end{array}\]
Validity in $\ATLSCPref^*$ is defined on state formulas wrt the empty strategy context as usual.

\section{Example: Translating Nash Equilibrium from $\ATLSCPref^*$ into $\QCTL^*$}
\label{exampleNash}

In this section we use the example of Nash equilibrium to illustrate the working of all the translations given in the previous sections. We show how the expression for Nash equilibrium in $\ATLSC^*$ with preference can be translated into an equivalent $\QCTL^*$ formula.

Consider our $\ATLSCPref^*$ formula 
\begin{equation}\label{NashEq}
\atlDclm{\Ag}\bigwedge\limits_{i\in\Ag}(\Next G_i\wedge\forallSim{}_i\mathbf{c}(G_i<_{\exists\forall,i}\mathbf{c}\Rightarrow\atlBclm{i}\Next\neg\mathbf{c})),
\end{equation}
for Nash equilibrium. Here $G_i$ are the $\LTL$ formulas expressing the objectives that the equilibrium strategy profile is supposed to achieve, $i\in\Ag$. Let $\Ag\defeq\{1,2\}$, and let $B_{I,i,k}$, $i\in\Ag$, $k=1,2,3$, be some $\LTL$ formulas which specify the classes of preference-indiscernible paths from $R_M^{\infin,-}(w_I)$ where $w_I$ is the initial state in the considered GCM $M$. (Here ${}_I$ indicates that the formulas define preference indiscernibility wrt the initial state $w_I$, and ${}_i$ denotes a player.)  Then
\[M,<,w_I\models\bigwedge\limits_{k=1}^3\exists\Next B_{I,i,k}\wedge\forall\Next\bigg(B_{I,i,k}\Rightarrow\bigwedge\limits_{k'\not=k}\neg B_{I,i,k'}\bigg),\ i\in\{1,2\}\]
Let  
\[M,<,w_I\models B_{I,i,1}<_i B_{I,i,2}\wedge B_{I,i,2}<_i B_{I,i,3}.\]
Using the expression (\ref{forallPathElim}) for $\forallSim\mathbf{p}C$ being $\forallSim{}_i\mathbf{c}(G_i<_{\exists\forall,i}\mathbf{c}\Rightarrow\atlBclm{i}\Next\neg\mathbf{c})$, we obtain the following $\forallSim{}_i\mathbf{c}$-free equivalent of (\ref{NashEq}): 
\[\atlDclm{1,2}\bigwedge\limits_{i\in\{1,2\}}
\left(
\Next G_i\wedge
\bigwedge\limits_{{\scriptsize \begin{array}{l} B_{i,1}\in \Cl(B_{I,i,1})\\
B_{i,2}\in \Cl(B_{I,i,2})\\
B_{i,3}\in \Cl(B_{I,i,3})\end{array}}} 
\left(\begin{array}{l}
q_{B_{i,1},1}\wedge q_{B_{i,2},2}\wedge q_{B_{i,3},3}
\Rightarrow\\ 
\bigwedge\limits_{X\subseteq \{1,2,3\}}
\left(G_i<_{\exists\forall,i}\left(\bigvee\limits_{k\in X}B_{i,k}\right)\Rightarrow
\atlBclm{i}\Next\neg\bigg(\bigvee\limits_{k\in X}B_{i,k}\bigg)
\right)
\end{array}
\right)
\right).
\]
Since the reference state is $w_I$, and the conjunction of the form $q_{B_{i,1},1}\wedge q_{B_{i,2},2}\wedge q_{B_{i,3},3}$ which holds in $w_I$ is $q_{B_{I,i,1},1}\wedge q_{B_{I,i,2},2}\wedge q_{B_{I,i,3},3}$, the formula can be simplified to
\[\atlDclm{1,2}\bigwedge\limits_{i\in\{1,2\}}\bigg(\Next G_i\wedge
\bigwedge\limits_{X\subseteq\{1,2,3\}}
\bigg(G_i<_{\exists\forall,i}\bigg(\bigvee\limits_{k\in X}B_{I,i,k}\bigg)\Rightarrow\atlBclm{i}\Next\neg\bigg(\bigvee\limits_{k\in X}B_{I,i,k}\bigg)\bigg)\bigg).
\]
The conjuncts for $X=\emptyset$ and $X=\Ag=\{1,2,3\}$ in the above formula hold trivially. The transitivity of $<_i$ implies that, for non-empty $X$, $G_i<_{\exists\forall,i}\bigg(\bigvee\limits_{k\in X}B_{I,i,k}\bigg)$ is equivalent to $G_i<_{\exists\forall,i}B_{I,i,\min X}$. Hence the formula can be further simplified to
\[\atlDclm{1,2}\bigwedge\limits_{i\in\{1,2\}}(\Next G_i\wedge\bigwedge\limits_{k=2}^3(G_i<_{\exists\forall,i}B_{I,i,k}\Rightarrow\atlBclm{i}\Next\bigvee\limits_{k'\not=k}B_{I,i,k'})).\]
Now the elimination of $<_i$ gives  
\begin{equation}\label{prefEliminated}
\atlDclm{1,2}\bigwedge\limits_{i\in\{1,2\}}\bigg(\Next G_i\wedge\bigwedge\limits_{k=2}^3\bigg(\bigg(\bigvee\limits_{k'<k}\exists\Next (G_i\wedge B_{I,i,k'})\bigg)\Rightarrow\atlBclm{i}\Next\bigvee\limits_{k'\not= k}B_{I,i,k'}\bigg)\bigg).
\end{equation}
It remains to eliminate $\atlDclm{1,2}$, which is done using the special case of the clause about $t_{i_1,\ldots,i_n}^{p_1,\ldots,p_n}(\atlDclm{\{i_k,\ldots,i_m\}}B)$ from Section \ref{theTranslationATLSCtoQCTL} for $n=0$, $\{i_k,\ldots,i_m\}\defeq\Ag=\{1,2\}$, and $B\defeq\bigwedge\limits_{i\in\{1,2\}}\bigg( \Next G_i\wedge\bigwedge\limits_{k=2}^3\bigg(\bigg(\bigvee\limits_{k'<k}\exists\Next (G_i\wedge B_{I,i,k'})\bigg)\Rightarrow\atlBclm{i}\Next\bigvee\limits_{k'\not= k}B_{I,i,k'}\bigg)\bigg)$:
\[t(\atlDclm{1,2}B)=\exists q_1\exists q_2\bigg(\bigwedge\limits_{i=1}^2\forall\Box\bigg(\bigvee\limits_{a\in\Act_i}\forall\Next(q_i\Leftrightarrow a)\bigg)\wedge
\forall\bigg(\Next\Box\bigg(\bigwedge\limits_{j=1}^2 q_j\bigg)\Rightarrow t_{1,2}^{q_1,q_2}(H)\bigg)
\bigg)\]
where 
\[
t_{1,2}^{q_1,q_2}(H)=\bigwedge\limits_{i\in\{1,2\}}\bigg(\Next G_i\wedge\bigwedge\limits_{k=2}^3\bigg(\bigg(\bigvee\limits_{k'<k}\exists\Next (G_i\wedge B_{I,i,k'})\bigg)\Rightarrow\neg t_{1,2}^{q_1,q_2}\bigg(\atlDclm{i}\neg\Next\bigvee\limits_{k'\not= k}B_{I,i,k'}\bigg)\bigg)\bigg).
\]
Finally
\[\begin{array}{l}
t_{1,2}^{q_1,q_2}\bigg(\atlDclm{1}\neg\Next\bigvee\limits_{k'\not= k}B_{I,1,k'}\bigg)=
\exists r\bigg(\forall\Box(\bigvee\limits_{a\in\Act_1}\forall\Next(r\Leftrightarrow a))\wedge
\forall\bigg(\Next\Box(q_2\wedge r)\Rightarrow t_{2,1}^{q_2,r}\bigg(\Next\bigvee\limits_{k'\not= k}B_{I,1,k'}\bigg)\bigg)\bigg)
\\
t_{1,2}^{q_1,q_2}\bigg(\atlDclm{2}\neg\Next\bigvee\limits_{k'\not= k}B_{I,2,k'}\bigg)=
\exists r\bigg(\forall\Box(\bigvee\limits_{a\in\Act_1}\forall\Next(r\Leftrightarrow a))\wedge
\forall\bigg(\Next\Box(q_1\wedge r)\Rightarrow t_{1,2}^{q_1,r}\bigg(\Next\bigvee\limits_{k'\not= k}B_{I,2,k'}\bigg)\bigg)\bigg)
\end{array}\]
and $t_{3-i,i}^{q_{3-i},r}(\Next\bigvee\limits_{k'\not= k}B_{I,i,k'})= \Next\bigvee\limits_{k'\not= k}B_{I,i,k'}$, $i=1,2$, because no game-theoretic operators occur in these formulas. This concludes the reduction of (\ref{NashEq}) to a $\QCTL^*$ formula.

The elimination of $\forallSim{}_i{\mathbf c}$ and $<_{\exists\forall,i}$ above is achieved upon reaching (\ref{prefEliminated}); the rest of the transformations are about moving to a $\QCTL^*$ formula by expressing the occurrences of the game-theoretic operator. Here follows what (\ref{prefEliminated}), which is a purely $\ATLSC^*$ formula, states in detail. After applying Modus Tolens to the implication in (\ref{prefEliminated}) we obtain:
\[\atlDclm{1,2}\bigwedge\limits_{i\in\{1,2\}}
\bigg(
\Next G_i\wedge
\bigwedge\limits_{k=2}^3
  \bigg(\atlDclm{i}\Next B_{I,i,k}
        \Rightarrow
        \bigg(\bigwedge\limits_{k'<k}\forall\Next \neg (G_i\wedge B_{I,i,k'})\bigg)         
  \bigg)
\bigg).
\]
Since $B_{I,i,1},B_{I,i,2},B_{I,i,3}$ is a full system, $\bigwedge\limits_{k'<k}\forall\Next \neg (G_i\wedge B_{I,i,k'})$ can be written as $\forall\Next\bigg(G_i\Rightarrow\bigvee\limits_{k'\geq k}B_{I,i,k'}\bigg)$. Hence we obtain
\begin{equation}\label{prefEliminatedSpec}
\atlDclm{1,2}\bigwedge\limits_{i\in\{1,2\}}
\bigg(
\Next G_i\wedge
\bigwedge\limits_{k=2}^3
  \bigg(\atlDclm{1}\Next B_{I,i,k}
        \Rightarrow
        \forall\Next\bigg(G_i\Rightarrow\bigvee\limits_{k'\geq k}B_{I,i,k'}\bigg)         
  \bigg)
\bigg).
\end{equation}
This formula states that a global (equilibrium) strategy profile exists such that objectives $G_1$ and $G_2$ are achieved and, furthermore, if one of the players can deviate in a way that lets it achieve $B_{I,i,k}$, then the respective $G_i$, which is achievable by sticking to the equilibrium profile, consists of plays that are as good as those from $B_{I,i,k}$ or better already. This statement is relevant for $k=2$ or $k=3$. Deviating to achieve $B_{I,i,1}$ is uninteresting because the $B_{I,i,1}$-plays are the least desirable ones. Given the meaning of $B_{I,i,1},B_{I,i,2},B_{I,i,3}$, the equilibria of interest are those about $G_i$ being $B_{I,i,2}\vee B_{I,i,3}$, $i=1,2$, because player $i$ has no incentive to deviate, if the equilibrium strategy already guarantees it $B_{I,i,3}$. Let's focus on player $1$. Let $H_2\defeq 
\Next G_2\wedge
\bigwedge\limits_{k=2}^3
  \bigg(\atlDclm{2}\Next B_{I,2,k}
        \Rightarrow
        \forall\Next\bigg(G_2\Rightarrow\bigvee\limits_{k'\geq k}B_{I,2,k'}\bigg)         
  \bigg)$ for the sake of brevity.

The instance of (\ref{prefEliminatedSpec}) for $G_1$ being $B_{I,1,2}\vee B_{I,1,3}$ is 
\[\atlDclm{1,2}
\bigg(
\Next (B_{I,1,2}\vee B_{I,1,3})\wedge
\bigwedge\limits_{k=2}^3
  \bigg(\atlDclm{i}\Next B_{I,1,k}
        \Rightarrow
        \forall\Next\bigg((B_{I,1,2}\vee B_{I,1,3})\Rightarrow\bigvee\limits_{k'\geq k}B_{I,1,k'}\bigg)         
  \bigg)
\wedge H_2\bigg).
\]
Observe that $\forall\Next((B_{I,1,2}\vee B_{I,1,3})\Rightarrow(B_{I,1,2}\vee B_{I,1,3}))$ is trivially true. Furthermore $\forall\Next((B_{I,1,2}\vee B_{I,1,3})\Rightarrow B_{I,1,3})$ is false because the class of $B_{I,1,2}$-plays is nonempty. Hence we obtain
\[\atlDclm{1,2}(\Next (B_{I,1,2}\vee B_{I,1,3})\wedge\neg \atlDclm{1}\Next B_{I,1,3}\wedge H_2).\]
In words, there exists an (equilibrium) strategy profile where player $1$ achieves $B_{I,1,2}\vee B_{I,1,3}$ and deviating cannot improve $1$'s score to just $B_{I,1,3}$.

\section*{Concluding Remarks}

In this paper we have proposed enhancing the language and semantics of $\ATLSC^*$ with a binary operator of preference and a specialized quantifier which binds variables with the status of path formulas that range over the sets of plays that are closed under preference indiscernibility. None of these additions to the language of $\ATLSC^*$ is completely unknown in the literature on its own right. The preference operator has been studied in both non-temporal settings \cite{DBLP:conf/kramas/DegremontK08,vanOtterloo2005preference} and in a temporal setting such as that of \cite{DBLP:journals/jolli/BaskentM20}. Quantification with similar restrictions, yet wrt epistemic indiscernibility and sets of states as opposed to preference-indiscernibility and the sets of plays we consider, has been proposed in \cite{DBLP:journals/tocl/BerthonMMRV21}. Our main contribution is to show that, under a reasonable assumption, these features of the language can be made available to enable formulating interesting properties of infinite multiplayer games in a way that is more convenient without the need to develop model checking for the extended system of $\ATLSC^*$ we obtain from scratch: both preference and the specialized quantifier can be eliminated and a translation into $\QCTL^*$, the established intermediate notation for automated reasoning used in so many logics of strategic ability now, can be used in our setting too. Among other things, we have proposed a new satisfaction preserving translation from $\ATLSC^*$ to $\QCTL^*$ and discussed its computational complexity.

\section*{Acknowledgements} The author thanks Antonio Cau for his helpful corrections and comments on this paper.


\newcommand{\etalchar}[1]{$^{#1}$}

\end{document}